\documentclass[ejsv2,noshowframe]{imsart}

\RequirePackage[numbers]{natbib}
\RequirePackage[colorlinks,citecolor=blue,urlcolor=blue]{hyperref}
\RequirePackage{graphicx}

\RequirePackage{tikz}
\RequirePackage{booktabs}
\usepackage{color}

\def\indep{\bot\!\!\!\bot}

\newcommand*{\colorboxed}{}
\def\colorboxed#1#{%
  \colorboxedAux{#1}%
}
\newcommand*{\colorboxedAux}[3]{%
  \begingroup
    \colorlet{cb@saved}{.}%
    \color#1{#2}%
    \boxed{%
      \color{cb@saved}%
      #3%
    }%
  \endgroup
}

\usepackage[T1]{fontenc}
\usepackage{newtxtext,newtxmath}

\startlocaldefs
\theoremstyle{plain}

\newtheorem{theorem}{Theorem}[section]
\newtheorem{lemma}[theorem]{Lemma}
\newtheorem{remark}{Remark}
\theoremstyle{definition}

\newtheorem{corollary}[theorem]{Corollary}
\newtheorem{example}{Example}

\def\T{ \mathrm{\scriptscriptstyle T} }
\theoremstyle{remark}

\endlocaldefs
\newtheorem{assumption}{Assumption}

\begin{document}
\begin{frontmatter}
\title{Mediation analysis with unmeasured confounding between parallel mediators and outcome}
\runtitle{Mediation analysis with unmeasured confounding}

\begin{aug}
\author[A]{\fnms{Kang}~\snm{Shuai}\ead[label=e1]{shuaikangpku@gmail.com}\orcid{0009-0007-3194-6930}},
\author[B]{\fnms{Lan}~\snm{Liu}\ead[label=e2]{liulan1815@outlook.com}},
\author[A]{\fnms{Yangbo}~\snm{He}\ead[label=e3]{heyb@math.pku.edu.cn}}
\and
\author[C]{\fnms{Wei}~\snm{Li}\ead[label=e4]{weilistat@ruc.edu.cn}}
\address[A]{School of Mathematical Sciences,
 Peking University\printead[presep={,\ }]{e1,e3}}

\address[B]{The Beijing Institute of Brain Disorders, Chinese Institutes for Medical Research, Capital Medical University\printead[presep={,\ }]{e2}}

\address[C]{Center for Applied Statistics and School of Statistics,
Renmin University of China\printead[presep={,\ }]{e4}}
\runauthor{Shuai et al.}
\end{aug}

\begin{abstract}
Mediation analysis extending beyond single mediators has gained significant attention in recent years. However, related methods often assume the absence of unmeasured mediator-outcome confounding. To address this, we develop a mediation analysis framework that accounts for such confounding within a linear structural equation model with parallel mediators. Specifically, we introduce a pseudo proxy variable to capture unmeasured confounding, allowing us to identify causal parameters. Leveraging this proxy, we propose a partially penalized method to identify mediators that significantly affect the outcome. The resultant estimates are consistent, and the estimates of nonzero parameters are asymptotically normal. Motivated by these results, we further introduce a procedure that can consistently select active mediation pathways with large probability. Simulation studies demonstrate the superior performance of the proposed approach. Finally, we apply our approach to genomic data, identifying gene expressions that potentially mediate the impact of a genetic variant on mouse obesity.
\end{abstract}



\end{frontmatter}

\section{Introduction}
Mediation analysis aims to explore how the effect of a treatment on an outcome is transmitted through a mediator variable. The primary objective is to disentangle the overall causal effect into a direct treatment-outcome link and an indirect effect through the mediator. Initially, mediation analysis focused on a single mediator within linear structural equation models \citep{baron1986moderator}. {This framework has since been extended to semiparametric methods (e.g., \citealp{tchetgen2012semiparametric}) and, more recently, to machine-learning approaches that enable flexible estimation of nonlinear mediation effects \citep{xu2022deepmed}. Meanwhile, the increasing availability of high-dimensional data has driven methodological advances extending mediation analysis to settings with multiple or high-dimensional mediators \citep{imai2013identification,vanderweele2014mediation,daniel2015causal,huang2016hypothesis,yang2025causal}. 

Mediation analysis with  numerous mediators has received considerable attention.  Within the linear structural equation modeling framework, various methods that were introduced in high-dimensional statistics have been adapted here for mediation analysis; see for instance, marginal screening, penalized regression, dimension reduction method, and  procedures that control the family-wise error rate or false discovery rate through multiple testing~\citep{zhang2016estimating,huang2016hypothesis,zhao2022pathway, sampson2018fwer, djordjilovic2019global,liu2022large}.  These approaches typically require no unmeasured confounding between mediators and the outcome. Neglecting  unmeasured confounders not only introduces bias in estimating direct and indirect effects, but may also lead to improper selections of crucial mediators in the analysis.} In the presence of unmeasured confounding, 
\citet{zheng2015causal} extended the framework of \citet{ten2007causal} to accommodate multilevel treatments and multicomponent mediators, requiring each mediator model to include covariate-treatment interactions. \citet{wickramarachchi2023mediation} built upon the work of \citet{fulcher2019estimation} to handle multiple mediators, exploiting heterogeneity in treatment effects. These identification methods require each mediator model to satisfy certain restrictions for addressing unmeasured confounding. While effective, they do not take advantage of parallel mediator structures that may be present in practice and could provide enhanced capabilities in addressing unmeasured confounding.

Many applied studies have adopted parallel mediator structures to simplify inference and enhance the interpretability of total effect decomposition, as each individual indirect effect can then be attributed to a specific mediator \citep{liu2013epigenome, lange2014assessing, huang2016hypothesis, taguri2018causal, mitt2019, jerolon2020causal}. 
 For example, \citet{liu2013epigenome} used epigenome-wide association studies to identify differentially methylated positions mediating genetic risk effects on rheumatoid arthritis. \citet{lange2014assessing} proposed a general framework for quantifying and ranking causal pathways through multiple mediators, while \citet{huang2016hypothesis} adopted a transformed mediation model to study microRNA miR-233’s role in Glioblastoma multiforme survival, modeling gene expression mediators as undirectedly correlated rather than sequentially ordered. \citet{zhang2016estimating} developed a joint significance test for mediation and applied it to assess DNA methylation markers as parallel mediators in the smoking--lung function pathway. Similarly, \citet{taguri2018causal} examined the impact of fluoride therapy on dental caries, considering oral bacteria levels and salivary fluoride as parallel mediators.

More recently, researchers have begun to leverage the shared confounding structure among parallel mediators to develop novel deconfounding techniques based on latent variable methods. This idea is inspired by  recent advances in causal inference with multiple treatments, which use shared confounding to identify average causal effects \citep{wang2019blessings, miao2023identifying, tang2023synthetic}. In the context of numerous mediators, \cite{derkach2019high} proposed a latent variable model for mediation analysis; however, their model assumes
that potential mediators are a group of latent factors, which differs from the settings
considered in this paper. \citet{yuan2024confounding} developed a latent factor model for parallel mediators  and established identification of average causal mediation effects under a latent sequential ignorability assumption, which requires the existence of a constructed surrogate confounder. Besides,
their method does not address the  mediator selection issue, and directly extending their framework to incorporate penalization may not be appropriate, as their model does not naturally accommodate regularization.

In this paper, we develop a mediation framework in the presence of shared unmeasured confounding between parallel mediators and the outcome. Specifically, we introduce a linear outcome model and a latent factor model for parallel mediators after excluding the effects of observed treatment and covariates. A crucial aspect of our analysis involves calculating the linear projection of the unmeasured confounder on the residual of regressing the parallel mediators on the treatment and observed covariates. The constructed projection variable can be seen as a pseudo proxy variable for the unmeasured confounding, enabling us to address both the mediation effect identification and the mediator selection challenge without external proxies. We replace the unmeasured confounder with the pseudo proxy variable within the outcome model and develop an adaptive lasso type procedure for estimation. In situations involving a univariate unmeasured confounder, our approach enables identification when only one mediator model contains nonlinear terms of baseline covariates or interactions between treatment and covariates. This contrasts with previous methods that require each mediator model to meet such restrictions. We also demonstrate the selection consistency of the resulting estimates and the asymptotic normality of the estimates for nonzero parameters. Lastly, we propose a  procedure that can consistently select active mediation pathways with large probability.

\section{Notation, model and identification}\label{notation}
Suppose we have $n$ independent and identically distributed observations from a population of interest. For each observation $i$, let $Z_i$ denote a treatment variable, $Y_i$ a continuous outcome, and $M_{i\cdot} = ( M_{i1},\ldots, M_{ip} )^{\T}$ a vector of continuous mediator variables lying in the causal pathways between the treatment and the outcome. Moreover, let $X_i\in \mathbb{R}^q$ denote a vector of pre-treatment covariates, and $U_i\in \mathbb{R}^t$ a vector of  unmeasured confounders.  {Throughout the following, the subscript $i$ will be omitted unless needed to avoid ambiguity. For clarity, the observed variables are $Z$, $Y$, $M$, and $X$, while the latent variables are represented by $U$.}

{We make the stable unit treatment value assumption \citep{rubin1980randomization} and adopt the potential outcomes framework to formalize causal problems.  Let $M_j(z)$ and $Y(z)$ denote the potential values that the $j$th mediator $M_j$ and the outcome $Y$ would achieve if the treatment $Z$ were set to level $z$. Similarly, let $Y(z,m_j)$ denote the potential outcome by simultaneously setting  $Z$ to level $z$ and $M_j$ to  $m_j$. 
In contrast, we let $Y\{z,M_j(z')\}$ denote the potential outcome where we do not specify $M_j$ to be some specific level, but set it to the level that would be potentially achieved under assignment to level $z'$ of the treatment. Analogous definitions apply to the potential values $M(z)$, $Y(z,m)$, and $Y\{z,M(z') \}$. The average total  causal effect of $Z$ on $Y$ is defined as $\mathrm{TE}= E \{ Y(z)-Y(z') \}$. The average natural direct effect (NDE) captures the effect achieved under treatment levels $z$ and $z'$, while maintaining the mediator at the value attained under a fixed treatment level $z'$, i.e., $\mathrm{NDE}=E[ Y\{ z,M(z') \} - Y\{ z',M(z') \}]$. The average natural indirect effect (NIE) quantifies the average change in the outcome when mediator $M$ is set to the values attained  under treatment levels $z$ and $z'$, while fixing treatment $Z$ at level $z$, i.e., $\mathrm{NIE}= E [ Y\{ z,M(z) \} - Y\{ z,M(z') \} ]$. Similarly, the natural indirect effect through $M_j$ is defined as $\mathrm{NIE}_j = E[ Y\{ z,M_j(z) \} - Y\{ z,M_j(z') \} ] $. Throughout the paper, the consistency assumption helps connect the observed variables to potential outcomes \citep{robins1992identifiability,pearl2001direct}.
}

The sequential ignorability assumption \citep{imai2010general, vanderweele2014mediation} is commonly imposed to identify natural direct and indirect effects. However, this assumption is violated when unmeasured confounding is present. To accommodate such confounding, we introduce a latent sequential ignorability assumption that explicitly allows for the existence of latent variables $U$.
\begin{assumption}[Latent sequential ignorability]\label{ass:latent} For $z,z_1',\ldots,z_p'\in\{0,1\}$, we assume:
    (i) $ M(z) \indep Z\mid X, U$; (ii) $Y(z,m) \indep Z \mid X,U$; (iii) 
$ Y(z,m) \indep M\mid Z,X,U$; (iv) $Y(z,m) \indep M(\boldsymbol{z}')\mid X,U$, where $M(\boldsymbol{z}')=\{M_1(z_1'),\ldots,M_p(z_p')\}$. 
\end{assumption}
Assumption \ref{ass:latent} permits $U$ to confound the treatment–mediator, treatment–outcome, and mediator–outcome relationships, thereby generalizing the standard sequential ignorability framework.
Assessing natural effects with multiple mediators often requires that
the mediators are causally ordered so that certain path-specific effects can be identified \citep{daniel2015causal}. 
many studies have also  focused on the case where the mediators do not causally affect each other \citep{liu2013epigenome, lange2014assessing,zhang2016estimating,taguri2018causal}, as mentioned in our introduction part. 
 \begin{figure}[b]
    \centering
\begin{tikzpicture}[>=stealth, scale=1.5]
    \node[draw, circle] (Z) at (0.5, 1) {$Z$};
     \node[draw, circle] (M1) at (2, 2) {$M_1$};
    \node[draw, circle] (A) at (2, 1) {\vdots};
     \node[draw, circle] (Mp) at (2, 0) {$M_p$};
    \node[draw, circle] (Y) at (3.5, 1) {$Y$};
    \node[draw, circle, densely dashed] (U) at (5, 1) {$U$}; 
    \draw[->] (Z) -- (A);
    \draw[->] (Z) -- (M1);
    \draw[->] (Z) -- (Mp);
    \draw[->] (A) -- (Y);
    \draw[->] (U) edge[bend right] (A);
    \draw[->] (U) -- (Y);
    \draw[->] (Z) edge[bend right] (Y);
    \draw[->] (M1) -- (Y);
    \draw[->] (Mp) -- (Y);
    \draw[->] (U) edge[bend right] (M1);
    \draw[->] (U) edge[bend left] (Mp);
\end{tikzpicture}\vspace{2mm}
    \caption{Causal graph with a vector of parallel mediators $M=(M_1,\ldots,M_p)$ and unmeasured confounders $U$.
		}
		\label{fig:high-dimensional}
\end{figure}
 Similar examples are  considered in  studies with multiple treatments \citep{miao2023identifying,tang2023synthetic}, which are common in many  contemporary applications such as genetics, recommendation systems and neuroimaging studies.
 To simplify the analysis, we thus focus on the setting with parallel mediators confounded by unmeasured variables, as shown in Figure~\ref{fig:high-dimensional}.
 Extensive discussions about this structure can
	be found in \citet{yuan2024confounding} and references therein. {For any $j\in\{1,\ldots,p\}$, define $M_{-j}=(M_1,\ldots,M_{j-1},M_{j+1},\ldots,M_p)$. Similar to the definition of $Y(z,m_j)$, let $M_{-j}(z,m_j)$ denote the potential outcome of $M_{-j}$ obtained by setting $Z=z$ and $M_j=m_j$. 
    The following assumption formally characterizes the parallel-mediator structure using the potential value of the mediators. 
    \begin{assumption}[Parallel-mediator]\label{ass:parallel}
     For  $z,z'\in\{0,1\}$, we assume:   (i) $M_{-j}(z,m_j)=M_{-j}(z)$; (ii) $M_{-j}(z)\indep M_j(z')\mid X,U$.
    \end{assumption}
    }

 Let $g(Z,X) = E(M\mid Z,X) \in \mathbb{R}^p$ and $\check{M} = M - g(Z,X)$, which represent the regression mean and residual of the mediators on treatment and observed covariates, respectively. We extend \cite{baron1986moderator} and propose the following structural equation models involving unmeasured variables $U$:
\begin{align}
	Y &= \beta_0 + \beta_1 Z + \beta_2^{\T} M + \beta_3^{\T} X + \varphi^{\T} U + \eta, \label{eqn:modelforY}\\
	\check{M} &= \Gamma U + \varepsilon. \label{eqn:modelforM}
\end{align}
Here, $\beta_2 \in \mathbb{R}^p$, $\beta_3 \in \mathbb{R}^q$, $\varphi \in \mathbb{R}^t$, $\Gamma \in \mathbb{R}^{p \times t}$, and we assume {$E(\eta\mid Z,M,X,U) = E(\varepsilon\mid Z,X,U) = 0$.} Let $\beta = (\beta_0, \beta_1, \beta_2^\T, \beta_3^\T)^\T$ and $\xi = (\beta^\T, \varphi^\T)^\T$.  
Although the treatment $Z$ is assumed to be of one-dimension here, it is also allowed to be  multi-dimensional in the models. As  shown in Appendix \ref{app:nonlinear}, the proposed approach can be extended to include additional nonlinear or interaction terms in the outcome model. However, for the sake of clarity, we focus on the simple yet  commonly-used linear model throughout the paper.

Model~\eqref{eqn:modelforM} implicitly assumes $E(U\mid Z,X) = 0$, which can be further relaxed by allowing $X$ to influence $U$. Specifically, if $U = \Psi X + \tilde{U}$ and $E(\tilde{U}\mid Z,X) = 0$, then $\tilde{U}$ can replace $U$ in models~\eqref{eqn:modelforY} and~\eqref{eqn:modelforM}. {Although Assumption~1 allows the unmeasured confounder $U$ to affect both the treatment $Z$ and the mediator $M$, the mean-independence restriction $E(U\mid Z,X)=0$, together with the additive structure in model~\eqref{eqn:modelforM}, ensures that the treatment--mediator effect remains identifiable within our framework.}

Without loss of generality, we assume $E(U) = 0_t$, $\mathrm{cov}(U) = I_t$. 
As previously assumed, the parallel mediators do not affect each other, and hence
 $\Sigma_{\varepsilon}=\mathrm{cov}(\varepsilon)$ in \eqref{eqn:modelforM} is diagonal. This essentially implies a latent factor model in \eqref{eqn:modelforM}. Then
 the parameter $\Sigma_{\varepsilon}$ is identifiable and the factor loading matrix $\Gamma$ is identifiable up to some rotation under certain conditions~\citep{anderson1956statistical}. 
{
\begin{lemma}\label{lem:causal-to-stat}
    Under Assumptions \ref{ass:latent}--\ref{ass:parallel} and models~\eqref{eqn:modelforY}--\eqref{eqn:modelforM}, we have
\begin{align*}
	\mathrm{NDE}  &= \beta_1 (z-z'),\\ \mathrm{NIE} &= \beta_2^{\T} E \{ g(z,X) - g(z',X) \},\\ 
    \mathrm{NIE}_j &=  \beta_{2j} E \{ g_j(z,X) - g_j(z',X) \},
\end{align*}
where $g_j(z,X)$ represents the $j$th component of $g(z,X)$.
\end{lemma}} 
From Lemma \ref{lem:causal-to-stat}, we conclude that estimating the  direct and indirect effects hinges on estimating  $\beta_1$ and $\beta_2$ in the outcome model~\eqref{eqn:modelforY}.
The outcome model involves the unmeasured confounder $U$, which poses challenges in identifying and estimating model parameters. 
Directly applying ordinary least squares (OLS) to regress $Y$ on $Z,M,X$ will yield biased estimates due to the correlation between $U$ and $M$. To address this issue, we employ projection techniques to extract the correlated component from $U$, ensuring that the projection residual and $M$ are uncorrelated. 
Specifically, through $L_2$ projection, we can select $\Delta \in \mathbb{R}^{p \times t}$ such that
	$\mathrm{cov} ( U - \Delta^\T \check{M}, \check{M} ) = 0$,
and by model~\eqref{eqn:modelforM}, we find $\Delta = (\Gamma \Gamma^{\T} + \Sigma_{\varepsilon} )^{-1} \Gamma \in \mathbb{R}^{p \times t}$. Defining $L = \Delta^{\T} \check{M} \in \mathbb{R}^t$
  and $\psi = \varphi^{\T} (U - L) + \eta$, we can reformulate model~\eqref{eqn:modelforY} as:
\begin{align}\label{eqn:transmodelforY}
	Y = \beta_0 + \beta_1 Z + \beta_2^{\T} M + \beta_3^{\T} X  + \varphi^{\T} L + \psi, 
\end{align}
where, under our model assumptions, the new error term $\psi$ can be shown to be uncorrelated with $Z$, $M$, $X$, $L$. It is important to note that the original term $\varphi^\T U$ in the outcome model is replaced by $\varphi^\T L$, so $L$ can be regarded as a pseudo proxy for the unmeasured confounder $U$.

{Intuitively, this construction can be interpreted as a control-function-style adjustment, a strategy commonly used to address endogeneity or latent confounding \citep{wooldridge2015control,guo2016control}. Under the mediator model \eqref{eqn:modelforM}, the correlation between $M$ and $U$ arises through the component of $U$ that is linearly related to the centered mediators $\check M$. Projecting $U$ onto $\check M$ produces $L$, which captures this confounding component. Once $L$ is included in the outcome model, the remaining part of $U$ becomes orthogonal to the mediators under our model assumptions, so the resulting error term is uncorrelated with the predictors and does not introduce endogeneity. In this sense, adjusting for $L$ effectively accounts for the confounding effect of the unmeasured variable $U$, enabling identification and consistent estimation of both direct and indirect effects, even when the mediators and the outcome share unmeasured confounding. While the latent sequential ignorability assumption holds when conditioning on the true unmeasured confounder $U$, it generally does not hold when conditioning on the pseudo proxy variable $L$. This follows from the construction of the pseudo proxy variable, which projects the unmeasured confounder $U$ onto the linear space spanned by all observed variables except the outcome $Y$, and therefore the pseudo proxy variable does not in general recover all confounding information contained in $U$. In contrast, \citet{yuan2024confounding} assumed the existence of a surrogate confounder such that latent sequential ignorability is preserved when conditioning on this surrogate variable. Besides, their surrogate confounder is defined under a high-level structural condition, which is conceptually more abstract than the construction of the pseudo proxy variable in our approach.}
Based on~\eqref{eqn:transmodelforY}, we next discuss the identification  of model parameters.


\begin{theorem}\label{THM:IDENTIFICATION_1}
	The vector of parameters $\beta$ is identifiable, and $\varphi$ is identifiable up to some rotation if the following conditions hold:
	\begin{itemize}
		\item[(i)] after deleting any row of $\Gamma$, there remain two disjoint sub-matrices of full column rank;
		
		\item[(ii)] the matrix $H = E\{ (1,Z,M^{\T},X^{\T},L^{\T} )^{\T} ( 1,Z,M^{\T},X^{\T},L^{\T}) \}$ is of full rank.
	\end{itemize}
\end{theorem}    
Condition (i) in Theorem~\ref{THM:IDENTIFICATION_1} is a standard requirement for identification in factor analysis. This condition ensures $\Gamma$ to be identifiable up to some rotation. When $U\in \mathbb{R}^1$, this condition requires $\Gamma$ to be a vector containing at least three non-zero components, implying that $U$ must confound at least three parallel mediators. If $U\in \mathbb{R}^t$, the factor loading matrix $\Gamma$ must contain at least $2t+1$ rows, which implies that there are at least $2t+1$ parallel mediators,  
and condition~(i) further requires that each variable in $U$ should confound at least three different mediators. 
 Condition (ii) involves the constructed predictor $L=\Delta^{\T} \check{M}$, which, given a fixed value of $\check{M}$, is identifiable up to a rotation. Although $L$ is not completely identifiable, condition (ii) can be tested for any chosen rotation. The proof in Appendix \ref{proof_thm1} demonstrates that if condition (ii) holds for a specific rotation of $L$, it will also hold for any other rotation. 
 Thus, without loss of generality, we can fix the rotation for ease of exposition. 
 By the definition of $L$, it is clear that condition (ii) essentially requires the vector-valued function $g(Z,X)$ to contain some nonlinear terms of treatment and baseline covariates. Below we provide an example to illustrate our identification conditions.

 \begin{example} \label{exam:identification}
	Suppose $X,U\in \mathbb{R}^1, M = (M_1, M_2, M_3)^{\T}$.
We  consider the following models:
	\begin{align*}
		Y  &= Z + M_1 + M_2 + M_3 + X + U + \eta, \\
		M_1 &= Z + X + Z X +  U + \varepsilon_1,  \\
		M_2 &= Z + X + U + \varepsilon_2, \\
		M_3 &= Z + X + U + \varepsilon_3, 
	\end{align*}
	where each predictor and residual are of zero mean and unit variance, and the residuals are mutually independent.  In this context, $\Gamma = (1,1,1)^{\T}$, thus fulfilling condition $(i)$. Furthermore, we have
	$$
	\Delta = (\Sigma_{\varepsilon} + \Gamma \Gamma^{\T})^{-1} \Gamma =
	\dfrac{\Sigma_{\varepsilon}^{-1} \Gamma}{1 + \Gamma^{\T} \Sigma_{\varepsilon}^{-1} \Gamma} = \dfrac{\Gamma}{4}.
	$$
	So we obtain
	$
	L=\Delta^{\T} \check{M} = 
	(M_1 + M_2 + M_3 - 3 Z - 3X - Z X)/4$.
	Condition $(ii)$ is also met due to the interaction term $ZX$ within $L$. 
\end{example}
Example~\ref{exam:identification} illustrates that our approach only requires $M_1$ to include the interaction term $ZX$ for identification. In contrast, the key assumption for identification in \citet{zheng2015causal} requires the vector $\{ Z - E(Z\mid X), E(M\mid Z,X) - E(M\mid X) \}$  to be non-degenerate under our model setting. A vector of random variables $(V_1,\cdots,V_k)$ is considered to be non-degenerate if, for all $ \lambda_1,\cdots,\lambda_k \in \mathbb{R}^1$, the condition $E(\sum_{i=1}^k \lambda_i V_i )^2 = 0 \Leftrightarrow \lambda_1= \cdots =\lambda_k=0$ holds. In Example~\ref{exam:identification}, $\{ Z - E(Z\mid X), E(M\mid Z,X) - E(M\mid X) \} = \{ Z-E(Z\mid X) \}(1,1+X,1,1)$ is degenerate and the identification assumption by \citet{zheng2015causal} fails. Their assumption can be satisfied if all three mediators include distinct interaction terms between treatment and baseline covariates. Meanwhile, the heterogeneity condition of \citet{wickramarachchi2023mediation} requires $\mathrm{var}(M\mid Z=z,X=x)$ to vary with $z$. However, in this example, the conditional variance is a constant vector. When each  $M_j$ incorporates a heterogeneous residual term $\varepsilon_j Z$ within the corresponding model, the conditional variance will vary with respect to $z$.

Our identification conditions employ interaction or nonlinear terms in certain mediator models as instruments rather than auxiliary variables. This approach aligns with common practices within the mediation analysis literature \citep{ten2007causal,small2012mediation,zheng2015causal}. Example~\ref{exam:identification} shows potentials of multiple mediators for identification due to their shared-confounding structure.
Another example highlights the difference of our method from the null-treatment or sparsity assumptions in two related papers by \citet{miao2023identifying} and~\citet{tang2023synthetic}, which can be found in Appendix \ref{examp}.

\section{Estimation and inference}\label{estimation}

{In the previous section, we established identification results  of the outcome model parameters $\beta$ under the proposed framework. Building on these  results, we now present a partially penalized procedure for estimating the outcome model parameters when $\beta_2$ is assumed to be sparse.}
 Due to the curse of dimensionality, conducting nonparametric estimation of $g(Z,X)$ often becomes impractical, especially when dealing with a large number of covariates. We thus propose a semiparametric model $g(Z,X;\gamma)$ parameterized by a finite-dimensional vector $\gamma\in \mathbb{R}^k$. Let the operator $\hat{E}[\cdot]$ denote the sample averaging operator. For a matrix $V$, let $\mathrm{vec}(V)$ denote the vectorization of $V$ and $\mathrm{diag}(V)$ the vector consisting of the diagonal elements of $V$.

Denote $\check{M}(\gamma) = M - g(Z,X;\gamma)$ and $\nu = \mathrm{vec}[ \{ \gamma,\Gamma,\mathrm{diag}(\Sigma_{\varepsilon}) \} ]$.
Let $\hat{\gamma} ,\hat{\Gamma},\hat{\Sigma}_{\varepsilon}, \hat{\nu}$ be the   estimators of $\gamma,\Gamma,\Sigma_{\varepsilon},\nu$, respectively. The estimation procedure is summarized as follows:
\begin{itemize}
    \item[(1)] Solve the minimization problem $\mathop{\min}_{\gamma} \hat{E} \lVert \check{M}(\gamma) \rVert_2^2$ to obtain $\hat{\gamma}$ and denote $\check{M}(\hat{\gamma})= M -  g(Z,X;\hat{\gamma})$, where $\lVert \cdot \rVert_2$ represents the $L_2$ norm;
    
    \item[(2)] Implement factor analysis  on $\check{M}(\hat{\gamma})$ to obtain $\hat{\Gamma}$ and $\hat{\Sigma}_{\varepsilon}$, then obtain $\hat{\Delta} = (\hat{\Gamma} \hat{\Gamma}^\T +\hat{\Sigma}_{\varepsilon})^{-1} \hat{\Gamma}$ and  an estimator  $\hat{L} = \hat{\Delta}^\T \check{M}(\hat{\gamma})$ of the proxy variable $L$;

    \item[(3)]  Solve the following adaptive lasso problem to obtain $\hat{\beta}_{\mathrm{ad}}$:
    \begin{align*}
 (\hat{\beta}_{\mathrm{ad}},\hat{\varphi}_{\mathrm{ad}})=  \mathop{\arg\min}_{ \beta, \varphi} \hat{E} \big( Y - \beta_0 - \beta_1 Z - \beta_2^\T M - \beta_3^\T X - \varphi^\T \hat{L} \big)^2 +\dfrac{ \lambda_n}{n} \sum_{r=1}^p \hat{w}_r \lvert \beta_{2,r} \rvert,
    \end{align*}
    where $\hat{w} =\lvert \hat{\beta}_{\mathrm{in},2} \rvert^{-\delta}$ using an initial 
    $\sqrt{n}$-consistent 
    estimator $\hat{\beta}_{\mathrm{in},2}$ of $\beta_2$,  and
    $\delta,\lambda_n>0$ are tuning parameters.
\end{itemize}
Step (1) corresponds to solving $k$ estimating equations to obtain $\hat\gamma$. Since $E\{ \check{M}(\gamma)\mid Z,X \}=0$, it follows that $E \{ \check{M}(\gamma) G(Z,X) \} = 0$ for any function $G(\cdot)$. Particularly, the minimization problem $\mathop{\min}_{\gamma} \hat{E} \lVert \check{M}(\gamma) \rVert_2^2$ in step (1) corresponds to solving  the following estimating equations:
\begin{align}\label{eqn:ee-zeta}
\hat E\bigg\{ \check{M}(\gamma)^\T \dfrac{\partial  g(Z,X;\gamma)}{\partial \gamma^\T} \bigg\} = 0.
\end{align}

The estimation procedure for $\Gamma$ and $\Sigma_{\varepsilon}$ in step (2) is performed by maximizing the following normal likelihood function, assuming $U\sim N(0,I_t)$ and $\varepsilon \sim N(0,\Sigma_{\varepsilon})$~\citep{anderson1956statistical}:
$
l(\Gamma,\Sigma_{\varepsilon}) =  -\log \lvert \Sigma \rvert - \mathrm{tr}( T_n \Sigma^{-1} )$, 
where $T_n = \hat{E} \{ \check{M}(\gamma) \check{M}(\gamma)^\T \}, \Sigma = \Gamma \Gamma^\T + \Sigma_{\varepsilon}$ and $\mathrm{tr}(\cdot)$ represents the trace operator. It is worth noting that the normal assumption is not essential in factor analysis because we can treat the likelihood as a quasi-likelihood.
Maximizing the above likelihood function in step (2) is equivalent to solving \eqref{eqn:ee-zeta} and the following estimating equations:
\begin{align}\label{eqn:ee-mle}
\hat E\bigg[  \dfrac{\partial}{\partial \alpha} \big\{ \log \lvert \Sigma \rvert + \check{M}(\gamma)^\T \Sigma^{-1} \check{M}(\gamma) \big\}  \bigg] =0,
\end{align}
where $\alpha = \mathrm{vec}[ \{ \Gamma,\mathrm{diag}(\Sigma_{\varepsilon}) \} ]$. 
Different from the classical adaptive lasso problem \citep{zou2006adaptive}, the minimization problem in step (3) involves an estimated variable $\hat L$ for $L$ and partially penalizes $\beta_2$ rather than the entire parameter vector. This introduces complexity in the theoretical analysis, because it requires consideration of additional uncertainty when deriving the asymptotic results of the proposed estimator. {For reproducibility, the estimation procedure can be implemented using several R packages and functions. In Step~(1), the parameter $\gamma$ in the conditional mean model $E(M \mid Z, X) = g(Z, X; \gamma)$ can be estimated using standard regression tools in R, such as the \texttt{lm()} or \texttt{glm()} functions. In Step~(2), which involves factor analysis on the residualized mediators $\check{M}(\hat{\gamma})$, the analysis can be carried out using the \texttt{factor.analysis} function in the \texttt{cate} package \citep{wang2017confounder}. In Step~(3), the adaptive lasso problem can be solved using the \texttt{glmnet} package.}
\begin{remark}\label{remark:lowd}
    In the estimation procedure described above, penalization via adaptive lasso is primarily intended for settings with a large number of mediators, where variable selection is necessary to identify the subset of mediators that have a true effect on the outcome. As we establish in Theorem \ref{THM:AD_NORMAL} later in this section, the adaptive lasso estimator enjoys the oracle property: it consistently selects the true set of nonzero coefficients, and the estimates of these coefficients are asymptotically unbiased and asymptotically normal. Consequently, in settings with a large number of mediators, the bias introduced by penalization are asymptotically negligible.

For settings with a low to moderate number of mediators, penalization is not required. In such cases, the parameters in the 
$Y$-model can be estimated directly by OLS. Specifically, replacing the adaptive lasso estimates in the estimation procedure (3), the OLS estimates are given by
\begin{align*}
 (\hat{\beta}_{\mathrm{ols}},\hat{\varphi}_{\mathrm{ols}})=  \mathop{\arg\min}_{ \beta, \varphi} \hat{E} \big( Y - \beta_0 - \beta_1 Z - \beta_2^\T M - \beta_3^\T X - \varphi^\T \hat{L} \big)^2,
    \end{align*}
which provides unbiased estimates under standard regularity conditions. Therefore, the choice of using penalization should depend on the dimensionality of the mediator vector: adaptive lasso is recommended for problems with a large number of mediators, whereas OLS is sufficient for low- or moderate-dimensional settings.
\end{remark}


Theorem~\ref{THM:IDENTIFICATION_1} shows that $\beta$ is identifiable, although the factor loading $\Gamma$ is identifiable only up to a rotation matrix. To ensure identifiability of $\Gamma$, a second condition is commonly imposed \citep{anderson1956statistical,bai2012statistical}; that is, 
$\Gamma^\T \Sigma_{\varepsilon}^{-1} \Gamma$ is assumed to be diagonal, with distinct positive elements arranged in decreasing order.
For the convenience of theoretical analysis, we retain this second condition to fix the rotation matrix of $\Gamma$. The asymptotic distribution of the estimator of $\beta_2$ remains unaffected by the rotation matrix. In other words, we can replace $\Gamma$ with $\Gamma A$ for any orthogonal matrix $A$, and the conclusions presented in this section will remain valid. 
We define
\begin{align*}
	Q(S;\nu)=\bigg[\check{M}(\gamma)^\T \dfrac{\partial  g(Z,X;\gamma)}{\partial \gamma^\T}, \dfrac{\partial }{\partial \alpha^\T } \big\{ \check{M}(\gamma)^\T {\Sigma}^{-1} \check{M}(\gamma) +  \log \lvert \Sigma \rvert \big\} \bigg]^\T,
\end{align*}
where $S=(Z,M^\T,X^\T)^\T$. We can summarize the estimating equations in~\eqref{eqn:ee-zeta} and~\eqref{eqn:ee-mle}  as:
$	\hat E\{ Q(S;\nu) \}=0$.
The estimator $\hat{\nu}$ is derived by solving these equations. Under certain regularity conditions, $\hat{\nu}$ is $\sqrt{n}$-consistent for $\nu_0$ and asymptotically normal, where $\nu_0$ is the true value of $\nu$.

The initial
$\sqrt{n}$-consistent 
estimator $\hat{\beta}_{\mathrm{in},2}$ in step (3) can be computed using various methods, such as least squares or the lasso procedure. However, with potentially many predictors, the lasso-type estimator is often preferred, especially when the true parameter is assumed to be sparse.
In this context, 
we present an initial lasso estimator, denoted as $\hat{\xi}_{\mathrm{la}}$, for $\xi$. This estimator is derived by partially penalizing $\beta_2$ in the following problem:
$$
\hat{\xi}_{\mathrm{la}} = (\hat{\beta}_{\mathrm{la}}^\T, \hat{\varphi}_{\mathrm{la}}^\T )^\T = \mathop{\arg\min}_{\beta,\varphi}\hat{E} \big( Y - \beta_0 - \beta_1 Z - \beta_2^\T M - \beta_3^\T X - \varphi^\T \hat{L} \big)^2  +   \dfrac{ \lambda_n}{n} \sum_{r=1}^p \lvert \beta_{2,r} \rvert.
$$
Let $R = (1,Z,M^\T,X^\T,L^\T)^\T$, and $R_i$ represents the $i$th realization of $R$. We summarize the $\sqrt{n}$-consistency of $\hat{\xi}_{\mathrm{la}}$ in the following theorem.
\begin{theorem}\label{THM:NORMAL}
Suppose that $\lambda_n = o_p(\sqrt{n})$ and the following conditions are satisfied:
\begin{itemize}
    \item[$\mathrm{(i)}$] $n^{-1} \max_{1\leq i \leq n} R_i R_i^\T \to 0$, and ${C}_n = n^{-1}\sum_{i=1}^n R_i R_i^\T \to C$ for a positive definite matrix $C$;
    
    \item[$\mathrm{(ii)}$] For $ 1\leq i,j \leq t$, the following expectations exist:
    $$
    E\bigg\{ R \dfrac{\partial L_j(\nu_0)}{ \partial \nu^\T } \bigg\},\quad\text{and}\quad  E\bigg\{ \dfrac{\partial L_i(\nu_0)}{ \partial \nu }  \dfrac{\partial L_j(\nu_0)}{ \partial \nu^\T } \bigg\}; $$
    
    \item[$\mathrm{(iii)}$]   $E( K K^\T )$ exists, where 
    $$
    K =   E\bigg[  R \dfrac{\partial \{ \varphi^\T L(\nu_0) \} }{\partial \nu^\T} \bigg]  \bigg[ \dfrac{\partial E\{ Q(S;\nu_0) \} }{\partial \nu} \bigg]^{-1}  Q(S;\nu_0) + R \psi.
    $$
\end{itemize}
Then under additional regularity conditions provided in Appendix \ref{proof_thm2}, we have:
$$
\sqrt{n} \big( \hat{\xi}_{\mathrm{la}}  -\xi \big) \overset{d}{\longrightarrow}  N(0,\Sigma_\mathrm{la}),
$$
where 
 $ \Sigma_\mathrm{la} = C^{-1} E \big( K K^\T \big) C^{-1}$.
\end{theorem}

The uncertainty associated with $\hat{\xi}_{\mathrm{la}}$ in Theorem~\ref{THM:NORMAL} stems from two distinct sources. Firstly, it comes from the procedure employed to estimate the parameters of $\nu_0$, and secondly, it arises from the process of estimating $\beta$ through the partially penalized least squares method. If the true value $\nu_0$ or $L$ were known, the optimization problem would become a standard lasso problem with a partial penalization term. Consequently, the covariance matrix $\Sigma_{\mathrm{la}}$ would be simplified to $C^{-1} E \big( R \psi^2 R^\T \big) C^{-1}$.
This demonstrates that the first term within 
$K$ accounts for the uncertainty in estimation of the constructed proxy variable $L$, which is also the impact of unmeasured confounding $U$ on the estimation of $\beta$. 


As shown in Theorem~\ref{THM:IDENTIFICATION_1},  $\beta$ is  identifiable while $\varphi$ is only identifiable up to some rotation, which implies that the constructed predictor $L$, and therefore $R$, are not completely identifiable. This raises an important question of whether the asymptotic normality of $\hat{\beta}_{\mathrm{la}}$ in Theorem~\ref{THM:NORMAL} depends on the rotation of $R$.
In particular, suppose that $R$ is replaced by $PR$ with $P$ defined as follows:
$$
P = 
\begin{pmatrix}
  I_{p+q+2}  & 0 \\
  0  &  A
\end{pmatrix},
$$
and $A$ denotes a rotation matrix that satisfies $A A^\T = A^\T A = I_t$. In this situation, because $\varphi^\T L$ is identifiable, the variable $K$ in condition (iii) of Theorem~\ref{THM:NORMAL} will similarly be replaced by $P K$, leading to an asymptotic variance of $\sqrt{n} \big( \hat{\xi}_{\mathrm{la}} - \xi \big)$ represented by:
$$
\big( P  C P^\T \big)^{-1} \big\{ P E ( K K^\T ) P^\T \big\} \big( P  C P^\T \big)^{-1}
 =P \Sigma_{\mathrm{la}} P^\T=\begin{pmatrix}
 	\Sigma_{11}  & \Sigma_{12} A^\T \\
 	A \Sigma_{12}^\T  & A \Sigma_{22} A^\T
 \end{pmatrix},
$$
where $\Sigma_{ij}$'s $(i,j=1,2)$ are block matrices of $\Sigma_{\mathrm{la}}$.
It is thus evident that the asymptotic variance of $\hat{\varphi}_{\mathrm{la}}$ and its asymptotic covariance with $\hat{\beta}_{\mathrm{la}}$ may be influenced by the rotation of $R$. However, the asymptotic variance of $\hat{\beta}_{\mathrm{la}}$ remains unaffected by any rotation of $R$.

Building upon the $\sqrt{n}$-consistent lasso estimator, we proceed to construct an initial estimator with all non-zero elements for $\beta_2$, given by $\hat{\beta}_{\mathrm{in},2} = \hat{\beta}_{\mathrm{la},2} + n^{-1}$, and introduce the adaptive weight $\hat{w}=\lvert \hat{\beta}_{\mathrm{in},2} \rvert^{-\delta}$ for some $\delta>0$.
Let $\mathcal{A}$ be the set of indices corresponding to the non-zero elements of $\beta_2$, and $\hat{\mathcal{A}}_n$ represent the set of indices for non-zero elements of $\hat{\beta}_{\mathrm{ad},2}$; that is, $\mathcal{A}=\{j:\beta_{2j}\neq 0\}$ and $\hat{\mathcal{A}}_n=\{j:\hat{\beta}_{\mathrm{ad},2j}\neq 0\}$. We then define $\Tilde{\mathcal{A}}$ as the union of index sets of $\beta_0$, $\beta_1$, $\beta_{2,\mathcal{A}}$, $\beta_3$, and $\varphi$.

\begin{theorem}
\label{THM:AD_NORMAL}
   Suppose conditions in Theorem~\ref{THM:NORMAL} hold and $\lambda_n n^{(\delta-1)/2} \to \infty$.
    Then we have
   \begin{itemize}
       \item[$\mathrm{(i)}$] consistency in variable selection: $\lim_{n\to \infty} P(\hat{\mathcal{A}}_n = \mathcal{A})= 1$,
       \item[$\mathrm{(ii)}$] asymptotic normality: $ \sqrt{n} ( \hat{\xi}_{\mathrm{ad}, \Tilde{\mathcal{A}}} - \xi_{\Tilde{\mathcal{A}}} ) \overset{d}{\longrightarrow} N(0,\Sigma_{\mathrm{ad}})$,
   \end{itemize}
   where 
   $ \hat{\xi}_{\mathrm{ad}} = (\hat{\beta}_{\mathrm{ad}}^\T, \hat{\varphi}_{\mathrm{ad}}^\T)^\T, 
\Sigma_{\mathrm{ad}} = \Tilde{C}^{-1} E(K_{\Tilde{\mathcal{A}}} K_{\Tilde{\mathcal{A}}}^\T) \Tilde{C}^{-1}$, and $ \Tilde{C} = C_{\Tilde{\mathcal{A}},\Tilde{\mathcal{A}}}.
   $
\end{theorem}
Theorem~\ref{THM:AD_NORMAL} shows that the adaptive lasso estimator $\hat{\beta}_{\mathrm{ad}}$ enjoys the oracle property. Specifically, the estimator $\hat{\beta}_{\mathrm{ad}}$ successfully identifies the true nonzero elements of $\beta_2$
 with probability asymptotically approaching 1, and the joint asymptotic distribution of the estimator $\hat{\beta}_{\mathrm{ad}}$ and $ \hat{\varphi}_{\mathrm{ad}}$ is the same as if the true underlying subset model were given in advance. The optimal values of the tuning parameters are chosen through  cross-validation procedures.
We have previously demonstrated that the asymptotic variance of $\hat{\beta}_{\mathrm{la}}$ remains unaffected by the rotation of $R$. The same holds true for the adaptive lasso estimator $\hat{\beta}_{\mathrm{ad}}$, with $C$ and $K$ replaced by $\Tilde{C}$ and $K_{\Tilde{\mathcal{A}}}$ in its asymptotic variance.
The asymptotic variance $\Sigma_{\mathrm{ad}}$ can be estimated using the observed data. Specifically, we can directly estimate $C$ by employing the sample mean $C_n = \hat{E} ( R R^\T )$. Likewise, $E( K K^\T )$ can also be estimated using the sample mean, incorporating estimators of all relevant parameters. By incorporating these estimators along with the estimated index set $\hat{\mathcal{A}}_n$, we can construct a consistent estimate of the asymptotic variance for the adaptive lasso estimator. 

 Because $\mathrm{NDE} = \beta_1 (z - z' )$, the asymptotic normality of the estimator $\hat{\beta}_{\mathrm{ad},1}$ in Theorem~\ref{THM:AD_NORMAL} allows us to evaluate the significance of the natural direct effect. The indirect effect through the $j$th mediator, denoted as $\mathrm{NIE}_j$, is equal to $\beta_{2j} \lambda_j$, with $ \lambda_j = E\{ g_j(z,X;\gamma) - g_j(z',X;\gamma) \}$.
  Traditional approaches for selecting active mediation pathways often involve performing the composite null hypothesis for each mediator $H_{0j}^c$: $\beta_{2j} \lambda_j=0$, which is complicated in practice. 
 Leveraging the selection consistency of the adaptive lasso estimator $\hat{\beta}_{\mathrm{ad},2}$, we can simplify the process of identifying active mediation pathways. Specifically, Theorem~\ref{THM:AD_NORMAL} implies that,  the mediators truly affecting the outcome can be asymptotically selected due to the oracle property of $\hat{\beta}_{\mathrm{ad},2}$.
 To determine the active mediation pathways, one can subsequently test whether $ \lambda_j=0$ among the selected mediators to identify those also influenced by the treatment variable. For example, when $g(Z,X)=\gamma_1 Z + \gamma_2X+\gamma_3 X^2$, the term $\lambda_j = \gamma_{1j} ( z - z')$, and it suffices to test whether $\gamma_{1j}=0$ for $j\in\hat{\mathcal{A}}_n$ using the standard $t$-test method.  More generally, one can construct an estimator $ \hat{\lambda}_j = \hat{E} \{ g_j(z,X;\hat{\gamma}) - g_j(z',X;\hat{\gamma}) \} $ for $\lambda_j$, and subsequently calculate the corresponding $z$-score to test whether $\lambda_j=0$ for $j\in\hat{\mathcal{A}}_n$. We summarize the procedure for selecting active causal mediation pathways as follows: (1) obtain the index set $\hat{\mathcal{A}}_n$ from the nonzero elements of $\hat{\beta}_{\mathrm{ad},2}$; (2) for each $j\in \hat{\mathcal{A}}_n$, perform a hypothesis test $H_{0j}:\lambda_j=0$. {In particular, define the test statistic
$T_{n,j} = \sqrt{n}\hat{\lambda}_j /\widehat{\mathrm{se}}(\hat{\lambda}_j)$,
where $\widehat{\mathrm{se}}(\cdot)$ denotes the corresponding standard error, and let $p_{n,j}$ be the associated $p$-value calculated as $p_{n,j} = 2\{1- \Phi(\lvert T_{n,j} \rvert) \}$.
We reject $H_{0j}$ at significance level $\alpha_j$ and define the estimated active mediator set
$\hat{\mathcal{A}}_{\mathrm{act},n}
=
\{j\in\hat{\mathcal{A}}_n:\;p_{n,j}\le \alpha_j\}$.}

\begin{corollary}\label{COR}
For each $j \in \hat{\mathcal{A}}_n$,
we have
\begin{align*}
    &\lim_{n\to \infty} P(p_{n,j}\leq \alpha_j) = \alpha_j~\text{if $H^c_{0j}$ is true;}\\
    & \lim_{n\to\infty} P(p_{n,j}\leq \alpha_j) =1 ~\text{if $H^c_{0j}$ is false.}
\end{align*}
Additionally, denote $\mathcal{A}_{\mathrm{act}} \subseteq \mathcal{A}$ as the true active mediator set, then 
we have
	$$
	\lim_{n\to \infty} P(\hat{\mathcal{A}}_{\mathrm{act},n} = \mathcal{A}_{\mathrm{act}}) \geq 1 - \sum_{j\in \mathcal{A}\backslash \mathcal{A}_{\mathrm{act}}}
	\alpha_j.
	$$

\end{corollary}

{
In Corollary~\ref{COR}, we show that the $p$-value obtained from testing $\lambda_j=0$ can also be used to evaluate the composite product null hypothesis $H^c_{0j}$. Specifically, when the composite null $H^c_{0j}$ is true, the probability that the corresponding $p$-value is below the prespecified threshold $\alpha_j$ converges to $\alpha_j$. In contrast, when $H^c_{0j}$ is false, this probability converges to 1. Therefore, the testing procedure based on $\lambda_j$ provides a valid asymptotic decision rule for the composite product test.} The consistency of selecting mediation pathways hinges on controlling the probability of rejecting at least one true $H_j$, also known as the family-wise error rate. This error rate should not exceed the aggregate significance levels, denoted by $\sum_{j\in \mathcal{A}\backslash \mathcal{A}_{\mathrm{act}}} \alpha_j$.  The classical Bonferroni correction method achieves error control at a predetermined significance level $\alpha$ by setting $\alpha_j = \alpha/h$ for each individual test $H_j$, where $h=|\hat{\mathcal{A}}_n|$ denotes the number of selected mediators in the first step. However, given the potential conservatism of the Bonferroni correction when $h$ is large, one can also adopt more powerful multiple testing techniques, such as Holm's method, Hochberg’s method or the Benjamini-Hochberg procedure \citep{benjamini1995controlling}, as implemented in our application section.

\section{Simulation}\label{sec:simulation}
In this section, we conduct simulation studies to assess the finite-sample performance of the proposed estimator. We compare our approach with two naive penalized regression methods, the naive lasso and naive adaptive lasso, which do not account for unmeasured mediator-outcome confounding. We generate the outcome $Y$ and mediators $M\in \mathbb{R}^p$ based on the following models: 
\begin{align*}
	Y &= \beta_1 Z + \beta_2^\T M + \beta_3 X + \varphi U + \eta,  \\
	M &= \gamma_1 Z + \gamma_2 X + \gamma_3 \exp(X) + \Gamma U + \varepsilon.
\end{align*}

Here, $(Z,X,U,\varepsilon,\eta)$ are drawn from a multivariate normal distribution with zero mean and identity covariance matrix. We set $\beta_1 = \beta_3 = 1$, $\gamma_1=\gamma_2=( 1, 1, 1,1, \ldots ,1 )^\T$ and let
\begin{subequations}
	\begin{align*}
		\gamma_3 &= ( 0.5, 0.5, 0.5, \underbrace{0, \ldots, 0}_{p-3} )^\T, \quad  \Gamma = ( 1, \ldots , 1, \underbrace{0, \ldots, 0}_{p-10} )^\T,
	\end{align*}
\end{subequations}
where the first three elements of $\gamma_3$ is non-zero, meaning that the first three mediators contain the nonlinear term $\exp(X)$, and the first ten elements of $\Gamma$ is non-zero, meaning that the unmeasured variable $U$ confounds the first ten mediators. We consider two scenarios for $\beta_2$:
\begin{align*}
	\text{Scenario 1 \& 2: } \beta_2 = ( 1, 1, 1, 1, 1, \underbrace{0, \ldots, 0}_{p-5} )^\T \text { and } \beta_2=( 1, 1, 1, 1, 1, 0,\ldots, 0, \underbrace{1 ,\ldots, 1}_{15} )^\T.
\end{align*}
Scenario 1 is constructed to include five mediators that are both active and confounded, while scenario 2 incorporates  additional 15 active but unconfounded mediators. 
We vary the sample size $n$ across $\{300, 600, 1000\}$ and the dimension $p$ across $\{100, 200, 300\}$. {In this section, we present simulation results for scenario~1 only. Results for scenario~2 are reported in Appendix~\ref{sec:simulation-scenario2} and they exhibit patterns similar to those observed in scenario~1.}

 Given that the effect of $Z$ on each $M_j$ has been fixed at 1, an individual mediation pathway is  active whenever the effect of $M_j$ on $Y$ is non-zero. Thus, our focus lies in assessing  precision of estimating $\beta_2$ and the selection consistency. 
 We compute the mean squared error (MSE) of each estimator, which is the sum of mean squared errors across the coordinates of the estimator:
	$\mathrm{MSE} = \sum_{j=1}^p (\hat{\beta}_{2j} - \beta_{2j})^2$,
where $\hat{\beta}_{2j}$ denotes an estimator of $\beta_{2j}$ for $j=1,\ldots,p$. We also calculate the number of true positives (TP) and false positives (FP) in estimating $\beta_2$. The results of MSE, TP, and FP for all methods, averaged over 200 experiments, are presented in Table~\ref{tab:beta2} 
for scenario 1.

The results  in Table~\ref{tab:beta2} indicate a significantly smaller MSE for our approach compared to the naive lasso and naive adaptive lasso methods. Notably, as the unmeasured confounding strength $\varphi$ increases from 1 to 4, the advantage of our approach becomes more apparent in terms of MSE, particularly evident in larger sample sizes.
From the TP results, we find that all methods appear to correctly identify the mediators that genuinely affect the outcome across various scenarios. However, concerning FP, our approach exhibits superior performance compared to the other two. The naive lasso exhibits the highest FP among the three methods. For cases with weaker unmeasured confounding strength (i.e., $\varphi=1$), the naive adaptive lasso demonstrates similar performance to our approach, both yielding nearly zero FP. However, as $\varphi$ increases, the FP of our approach remains consistently lower than that of the naive adaptive lasso in all scenarios. Moreover, with larger sample sizes, the FP of our approach converges to approximately zero, whereas the naive adaptive lasso maintains an FP of around 5. This value aligns with the number of mediators confounded by $U$ yet not affecting the outcome. This distinction highlights that our approach can effectively filter out false signals arising from unmeasured confounding, while the other two methods might incorrectly identify them as genuine signals.

These simulation results demonstrate that our approach can successfully identify the mediators with significant effects on the outcome. It can also eliminate nearly all false signals when unmeasured mediator-outcome confounding is present. In contrast, the two naive methods often misinterpret mediators confounded by $U$ as valid signals.
 Specifically, for the more competitive naive adaptive lasso method in scenario 1, all the ten mediators confounded by $U$ are incorrectly identified as the true mediators, although half of them do not affect the outcome. Our approach can accurately differentiate between genuine signals and those that have spurious correlations with the outcome. 
 Despite the strong performance of all  methods in terms of TP in the current settings, the presence of confounding bias may also lead the naive approachs to miss certain valid signals, which will induce lower TP values in certain cases for the two naive approaches. {In addition, results for estimating NIE in scenario~1 are presented in Table~\ref{tab:NIE}, and their performance patterns are consistent with those observed for $\beta_2$ in Table~\ref{tab:beta2}. In particular, the TP and FP  for NIE are very similar to those for $\beta_2$. This is expected because $\beta_2$ represents the effect of the mediator on the outcome, while the NIE is the product of $\beta_2$ and the effect of the treatment on the mediator. In our simulation setting, there is no unmeasured confounding between the treatment and mediators, so the treatment–mediator effect can be accurately estimated using standard regression methods. As a result, the estimation performance for NIE closely mirrors that for $\beta_2$.

To further assess the inference performance of our procedure, we modify scenario~1 by setting the fifth component of $\gamma_1$ to zero, that is, $\gamma_1 = (1,1,1,1,0,1,\ldots,1)^\T$, while letting the confounding strength be $\beta_4 = 4$ and the mediator dimension $p = 100$. Under scenario 1, only the first five mediators have nonzero elements in $\beta_2$ and therefore exert effects on the outcome. In this setting, our method successfully identifies these five mediators, and Table~\ref{simu_mediate_pathway} reports the bias and standard error (SE) of the estimated natural indirect effects through them for sample sizes $n \in \{1000, 2000\}$.
In addition, Table~\ref{simu_mediate_pathway} presents the rejection probability $P(p_{n,j} \le 0.05)$ for testing whether the treatment has an indirect effect on the outcome through mediator $j$. Since $\gamma_1$ determines the treatment–mediator relationship, only the first four mediators are truly affected by the treatment, while the fifth mediator is not. Consistent with this data-generating mechanism, the first four mediators yield  probabilities essentially equal to~1, whereas the probability for the fifth mediator is around $0.045$, below the nominal significance level $0.05$. This result confirms that only the first four mediation pathways are active, which is fully aligned with the theoretical guarantee in Corollary~\ref{COR}.

 }

\begin{table}[t]
	\centering
	\caption{Simulation results of the mean squared error (MSE), the number of true positives (TP) and false negatives (FP) for estimating $\beta_2$ by the proposed approach, naive lasso and naive adaptive lasso in scenario 1.}
	\label{tab:beta2}
	\resizebox{0.95\columnwidth}{!}{
	\begin{tabular}{ccccccccccccc}
		\toprule
		& & \multicolumn{3}{c}{Proposed approach} & & \multicolumn{3}{c}{Naive lasso} & & \multicolumn{3}{c}{Naive adaptive lasso}\\
		\cmidrule(r){3-5} \cmidrule(r){7-9} \cmidrule(r){11-13}  
		$p$ &   & 100  & 200 & 300 & & 100  & 200 & 300 & & 100  & 200 & 300\\
		\midrule
		& MSE    & 0.04  & 0.03  & 0.04  &  & 0.12  &  0.11 & 0.12 & & 0.15  & 0.15  & 0.15 \\
		$n=300,\varphi = 1 $	& TP    & 5  &  5 & 5  &  & 5  & 5  &  5  & &  5  & 5  & 5 \\
		& FP    & 0.23  & 0.01  & 0.01  &  & 11.51  &  13.94 &  15.93 & &   0 &  0 & 0\\
		\midrule
		& MSE   & 0.20  & 0.29  & 0.10  &  & 1.40  & 1.37  & 1.38 & & 1.50 & 1.51  & 2.05  \\
		$n=300,\varphi = 4 $	& TP   & 5  & 5  & 5  &  & 5  & 5  & 5 & &  5  & 5  & 5 \\
		& FP   & 2.78  & 7.01  &  0.28 & & 11.24  & 13.29  & 15.66 & &   5.27 & 5.57  & 1.37 \\
		\midrule
		& MSE    & 0.02  & 0.02  & 0.02  &  & 0.10  & 0.10 & 0.10 &  & 0.14  & 0.14  &  0.14 \\
		$n=600,\varphi = 1 $	& TP    & 5  & 5  & 5 &  &  5 & 5  & 5 & &  5 & 5  & 5 \\
		& FP   & 0  & 0   & 0  &  & 12.55  & 14.09  & 15.25 & & 0  &  0  & 0\\
		\midrule
		& MSE  & 0.05  & 0.04  & 0.04  &  & 1.37  &  1.37 &  1.36 & & 1.41 & 1.42  & 1.42\\
		$n=600,\varphi = 4 $	& TP   & 5  & 5  & 5  &  & 5  & 5  & 5 & & 5 & 5  & 5 \\
		& FP    & 0.06  & 0  &  0 &  & 11.41  & 13.82  & 15.03 & &   5 & 5  & 5 \\
		\midrule
		& MSE    & 0.01  & 0.01  & 0.01  & & 0.09  & 0.09  & 0.09 & & 
		0.13  & 0.13  &  0.13 \\
		$n=1000,\varphi = 1 $	& TP    & 5  & 5  & 5 &  & 5  & 5   & 5  & & 5  &  5 &  5 \\
		& FP   & 0  & 0  & 0  &   & 11.53 & 14.80  & 16.07  & & 0  & 0  &  0     \\
		\midrule
		& MSE     & 0.03  & 0.03  & 0.03  & & 1.37  & 1.36  & 1.37 &  & 1.40  & 1.39  &  1.39\\
		$n=1000,\varphi = 4 $ & TP   & 5  & 5  & 5 &  & 5  & 5   & 5 & & 5  & 5  & 5 \\
		& FP  & 0.01 & 0  & 0 &   & 11.31 & 13.69  & 15.15 & & 5  & 5  & 5  \\
		\bottomrule
	\end{tabular}%
	}
\end{table}%

 \begin{table}[t]
	\centering
	\caption{{Simulation results of the mean squared error (MSE), the number of true positives (TP) and false negatives (FP) for estimating $\mathrm{NIE}$ by the proposed approach, naive lasso and naive adaptive lasso in scenario 1.}}
	\label{tab:NIE}
	\resizebox{0.95\columnwidth}{!}{
	\begin{tabular}{ccccccccccccc}
		\toprule
		& & \multicolumn{3}{c}{Proposed approach} & & \multicolumn{3}{c}{Naive lasso} & & \multicolumn{3}{c}{Naive adaptive lasso}\\
		\cmidrule(r){3-5} \cmidrule(r){7-9} \cmidrule(r){11-13}  
		$p$ &   & 100  & 200 & 300 & & 100  & 200 & 300 & & 100  & 200 & 300\\
		\midrule
		& MSE    & 0.18 & 0.15  & 0.14  &  & 0.28  &  0.25 & 0.27 & & 0.30  & 0.31  & 0.32 \\
		$n=300,\varphi = 1 $	& TP    & 5 & 5  & 5  &  & 5  &  5 & 5 & & 5  & 5  & 5 \\
		& FP    & 0.23 & 0.01  & 0.01  &  & 11.51  &  13.94 & 15.93 & & 0  & 0  & 0\\
		\midrule
		& MSE   & 0.35 & 0.42  & 0.22 &  & 1.67  &  1.58 & 1.66 & & 1.69  & 1.70  & 2.36  \\
		$n=300,\varphi = 4 $	& TP   & 5 & 5  & 5  &  & 5  &  5 & 5 & & 5  & 5  & 5 \\
		& FP   & 2.78 & 7.01  & 0.28  &  & 11.24  &  13.29 & 15.66 & & 5.27  & 5.57  & 1.37 \\
		\midrule
		& MSE    & 0.08 & 0.07  & 0.08  &  & 0.17  &  0.17 & 0.17 & & 0.22  & 0.24  & 0.23 \\
		$n=600,\varphi = 1 $	& TP    & 5 & 5  & 5  &  & 5  &  5 & 5 & & 5  & 5  & 5 \\
		& FP   & 0 & 0  & 0  &  & 12.55  &  14.09 & 15.25 & & 0  & 0  & 0\\
		\midrule
		& MSE  & 0.11 & 0.10  & 0.11  &  & 1.47  &  1.50 & 1.47 & & 1.51  & 1.59  & 1.56\\
		$n=600,\varphi = 4 $	& TP   & 5 & 5  & 5  &  & 5  &  5 & 5 & & 5  & 5  & 5 \\
		& FP    & 0.06 & 0  & 0  &  & 11.41  &  13.82 & 15.03 & & 5  & 4.99  & 5 \\
		\midrule
		& MSE    & 0.05 & 0.05  & 0.05  &  & 0.14  &  0.15 & 0.13 & & 0.20  & 0.20  & 0.18 \\
		$n=1000,\varphi = 1 $	& TP    & 5 & 5  & 5  &  & 5  &  5 & 5 & & 5  & 5  & 5 \\
		& FP   & 0 & 0  & 0  &  & 11.53  &  14.80 &16.07 & & 0  & 0  & 0     \\
		\midrule
		& MSE     & 0.07 & 0.07  & 0.07  &  & 1.45  &  1.48 & 1.42 & & 1.53  & 1.53  & 1.46\\
		$n=1000,\varphi = 4 $ & TP   & 5 & 5  & 5  &  & 5  &  5 & 5 & & 5  & 5  & 5 \\
		& FP & 0.01 & 0  & 0  &  & 11.31  &  13.69 & 15.15 & & 5  & 5  & 5  \\
		\bottomrule
	\end{tabular}%
	}
\end{table}%

\begin{table}[t]
	\centering
	\setlength{\abovecaptionskip}{0.cm}
	\caption{{Bias and standard error (SE) of the natural indirect effects through the selected mediators, as well as the rejection probability $P(p_{n,j}\le 0.05)$ for testing whether the treatment affects the outcome through mediator $j$, where $p_{n,j}$ denotes the corresponding $p$-value.
    The mediators framed in red are found to be active in the corresponding mediation pathway.
    }}
	\label{simu_mediate_pathway}
	\setlength{\tabcolsep}{3mm}
	\resizebox{0.97\textwidth}{!}{
		\begin{tabular}{ccccccccc}
			\toprule
            \multicolumn{4}{c}{$n=1000$} & &  \multicolumn{4}{c}{$n=2000$}\\
            \midrule
     	Mediator index $j$ & Bias & SE  & $P(p_{n,j}\leq 0.05)$ & & Mediator index $j$ & Bias & SE  & $P(p_{n,j}\leq 0.05)$  \\
			\midrule
			\colorboxed{red}
            {1} &0.022	 &0.442	& 1 & & \colorboxed{red}{1}  &0.036 &0.414 &1 \\
                \colorboxed{red}{2}	 &0.023	 &0.446	&1 & & \colorboxed{red}{2} &0.032 &0.412 &1	 \\
                \colorboxed{red}{3}	 &0.025	 &0.440	&1 &  & \colorboxed{red}{3} &0.030 &0.412 &1 \\
                \colorboxed{red}{4}	 &0.051	 &0.482	&1 & & \colorboxed{red}{4} &0.062 &0.452 &1 \\
                5 &0.004  &0.271 &0.045	&& 5 &0.002 &0.254 &0.045  \\	
			\bottomrule
		\end{tabular}%
	}
\end{table}%

\begin{table}[t]
	\centering
	\setlength{\abovecaptionskip}{0.cm}
	\caption{Estimate and standard error (SE) of natural indirect effects through selected genes,  as well as the corresponding $p$-value of testing each mediation pathway by the proposed approach. The genes framed in red are found to be active by all multiple testing correction procedures.}
	\label{mediate_pathway}
	\setlength{\tabcolsep}{3mm}
	\resizebox{0.97\textwidth}{!}{
		\begin{tabular}{ccccccccc}
			\toprule
			Gene & Estimate & SE  & $p$-value & & Gene & Estimate & SE  & $p$-value   \\
			\midrule
			\colorboxed{red}{Slc39a11} &0.131	 &0.072	& $<10^{-3}$ & &Vkorc1l1	&-0.038	&0.124	&0.205 \\
                \colorboxed{red}{Ercc3}	 &0.144	 &0.272	&0.001	   &  &Lamc1	    &0.0144	&0.041	&0.252 \\
                Psmd12	 &0.124	 &0.118	&0.003	   &  &Serpina12	&-0.013	&0.027	&0.316 \\
                Zfp533	 &0.172	 &0.275	&0.005	    & &Tbx 22	&0.006	&0.118	&0.357 \\
                Serpina6 &-0.157 &0.094 &0.006	    & &Cmas	    &-0.019 &0.129 	&0.522 \\
                RGD1563955 &-0.139 &0.121 &0.012	& &Rn.20259	&0.035	&0.100	&0.536 \\
                Nek2	 &-0.143 &0.297	 &0.012	    & &Psmd8	    &-0.026 &0.140	&0.550 \\
                Slc43a1	 &0.043	 &0.058	&0.040	    & &Igsf5	    &-0.019	&0.105  &0.680 \\
                Adamts8	 &0.010	 &0.083	 &0.047	    & &Rn.8483	&0.003  &0.151  &0.702 \\
                Pigr	 &0.073	 &0.015 &0.049	    & &F2r	    &-0.009 &0.057	&0.812 \\
                S100a10	 &0.052	 &0.038	 &0.062	    & &Tsc22d1	&0.001	&0.029	&0.895 \\
                Trim13	 &0.074  &0.141	 &0.125	    & &Prdm1	    &-0.003 &0.080	&0.947 \\
                Agtr1a	 &0.034	 &0.072	 &0.169	    & &Avpr1a	&-0.004	&0.074	&0.954 \\		
			\bottomrule
		\end{tabular}%
	}
\end{table}%

 \begin{table}
	\centering
	\setlength{\abovecaptionskip}{0.cm}
	\caption{Analysis results of the mouse obesity data by the naive adaptive lasso method. The genes in bold show those that are selected only by this method, and
		the rest of the caption details
		remain the same as those in Table \ref{mediate_pathway}.
	}
	\label{mediate_pathway_naive}
	\setlength{\tabcolsep}{3mm}
	\resizebox{0.97\textwidth}{!}{
		\begin{tabular}{ccccccccc}
			\toprule
			Gene & Estimate & SE  & $p$-value & &Gene & Estimate & SE  & $p$-value   \\
			\midrule
			\colorboxed{red}{\textbf{Hip2}}	&0.076   &0.405	 &$<10^{-4}$ & &Vkorc1l1	&-0.048	 &0.107	&0.205 \\
               \colorboxed{red}{Slc39a11} &0.122	 &0.081  &$<10^{-3}$  & & \textbf{Dot1l}    &-0.002  &0.057	&0.205 \\
               \colorboxed{red}{Ercc3}	&0.167   &0.088	 &0.001      & &Lamc1	&0.012	 &0.029	&0.252 \\
                Psmd12	&0.113   &0.138  &0.003	     & &\textbf{BC022687}	&$<10^{-4}$ &0.029	&0.267 \\
                Zfp533	&0.169   &0.133	 &0.005      & &Serpina12 &-0.020 &0.057 &0.316  \\
                Serpina6 &-0.154 &0.066  &0.006	     & &Tbx 22	&0.004   &0.120 &0.357 \\
                RGD1563955 &-0.151 &0.132 &0.012	 & &Cmas	    &-0.023  &0.118 &0.522 \\
                Nek2	&-0.152  &0.082  &0.012      & &Rn.20259	&0.035  &0.101  &0.536 \\
                Slc43a1	&0.037   &0.033	 &0.040	     & &Psmd8	&-0.025 &0.148  &0.550 \\
                Adamts8	&0.023   &0.068  &0.047      & &Igsf5	&-0.018 &0.137  &0.680 \\
                Pigr	&0.099	 &0.114  &0.049	     & &Rn.8483	&0.004	&0.127  &0.702 \\
                \textbf{Smad5}	&0.006	 &0.032  &0.061	     & &F2r	    &-0.009 &0.055  &0.812 \\
                S100a10	&0.061	 &0.047  &0.062	     & &Tsc22d1	&0.002  &0.043  &0.895 \\
                Trim13	&0.071   &0.106  &0.125	     & &Prdm1	&-0.003 &0.083	&0.947 \\
                \textbf{Sh3yl1}	&0.005	 &0.038  &0.158	    &  &Avpr1a	&-0.004	&0.121	&0.954 \\
                Agtr1a	&0.029	 &0.038  &0.169       & &   &  &  & \\				
			\bottomrule
		\end{tabular}%
	}
\end{table}%

\section{Data analysis}\label{DATA_ANA}
In this section, we use the mouse obesity data described by~\citet{wang2006genetic} to illustrate the proposed approach. It focuses on
evaluating the effect of gene expressions on the body weight of F2 mice.  In this context, it is important to consider potential unmeasured phenotypes that might confound the gene expressions and body weight. \citet{lin2015regularization} employed single nucleotide polymorphisms (SNPs) as instrumental variables and introduced a high-dimensional instrumental variable regression approach for analyzing this dataset. However, the presence of potential pleiotropic effects poses a challenge, because the SNPs might violate the exclusion restriction assumption of instrumental variables. Among the pool of SNPs, a specific variable named rs4231406 was previously identified as a quantitative trait locus for atherosclerosis, with a significant association with both body weight and adiposity~\citep{wang2006genetic}. Here, our goal is to investigate the direct effect of this SNP on body weight as well as its potential indirect effect through gene expressions using mediation analysis. The dataset we use includes 287 mice and comprises two observed covariates: sex and SNP density.
We further incorporate the top 200 gene expressions that exhibit the strongest correlations with body weight as candidate mediators.

We apply the proposed approach to analyze this data and 
employ the naive adaptive lasso method for comparison.
The results show that the proposed approach estimates the natural direct effect to be 0.251, exhibiting a standard deviation of 0.443 and a corresponding $p$-value of 0.572. This suggests that the SNP identified as rs4231406 does not appear to significantly influence the outcome directly. The results obtained from the naive adaptive lasso method provide similar findings, with an estimated natural direct effect of 0.193, a standard deviation of 0.366, and a $p$-value of 0.598. 

Table~\ref{mediate_pathway} presents the results of the proposed approach, excluding 3 genes with missing information. The results for the naive adaptive lasso method is displayed in Table \ref{mediate_pathway_naive}. Both methods identify 26 genes that affect the body weight, while the naive method includes 5 more genes, namely Hip2, Smad5, Sh3yl1, Dot1l, and BC022687. These findings align with the simulation results, which indicate an increased number of false positives in the naive method. In Table~\ref{mediate_pathway}, we also present the calculated $p$-values for testing mediation pathways via the selected gene expressions. Given the presence of multiple testing challenges, we apply various correction methods like Bonferroni, Holm's, Hochberg's and Hommel's to control the family-wise error rate. These corrections confirm that our proposed approach identifies active mediation pathways through the genes Slc39a11 and Ercc3, while the naive adaptive lasso method selects an additional pathway through Hip2. {It is worth noting that \citet{liu2013zinc}  demonstrated a strong correlation between Slc39a11 and obesity in mice. In fact, the gene {Slc39a11} encodes a member of the ZIP (Zrt-/Irt-like) family of metal ion transporters, particularly involved in zinc transport. Zinc transporters are widely implicated in adipogenesis and adipocyte function~\citep{yu2013characterization}, suggesting that {Slc39a11} may affect body weight in mice. The gene {Ercc3} is a DNA helicase subunit of TFIIH, central to nucleotide excision repair~\citep{vermeulen1994clinical}, and may influence growth, development, endocrine function, and systemic metabolism. Natural or regulatory variation near {Ercc3} could therefore plausibly associate with baseline body weight.} In contrast, the additional pathway selected by the naive adaptive lasso involves Hip2, which is associated with E2 ubiquitin-conjugating enzymes and neurodegenerative diseases and may not significantly influence body weight in mice. This is likely attributable to the fact that Hip2 is primarily linked to neurodegenerative processes rather than obesity-related mechanisms \citep{su2018reduction}. Furthermore, employing the Benjamini-Hochberg procedure to control the false discovery rate reveals 5 more significant pathways via Psmd12, Zfp533, Serpina6, RGD1563955, and Nek2 for both methods. With more sample sizes, our approach is expected to be more reliable for  selecting active mediation pathways under unmeasured mediator-outcome confounding.

\section{Discussion}\label{discuss}
This paper presents an approach for  mediation analysis with unmeasured confounding between parallel mediators and the outcome. 
By constructing a pseudo proxy variable for unmeasured confounding, we introduce identification conditions and a partially penalized adaptive lasso procedure for estimation. We show that the resulting estimates are consistent and the estimates of nonzero parameters are asymptotically normal.
Then we propose a procedure to select active causal mediation pathways,
which consistently selects active mediation pathways with large probability.


 The proposed approach may be improved or extended in several directions. First, we can extend the identification results to accommodate scenarios involving interactions between the treatment, mediator, observed covariates, and even other complex functions of the treatment and covariates within the outcome model under  additional conditions akin to those outlined in Theorem~\ref{THM:IDENTIFICATION_1}(ii); see Appendix \ref{app:nonlinear} for more details.
  {Second,
  our theoretical results assume a fixed number of latent confounders, but this number does not need to be small. To ensure identification, the latent dimension $t$ is related to the number of mediators $p$, with a necessary condition for condition (i) in Theorem~\ref{THM:IDENTIFICATION_1} being $p \ge 2t + 1$. Consequently, a relatively large number of latent confounders can be accommodated when $p$ is  large as long as condition (i) holds. Assuming a predetermined latent dimension $t$ is common in factor analysis, as the number of factors can often be consistently estimated \citep{bai2002determining,wang2017confounder}. In practice, $t$ can be selected using established methods for factor models, such as the principal component–based criteria of \citet{bai2002determining} or the bi-cross-validation method of \citet{owen2016bicrossvalidation}. Currently, our identification strategy assumes linear dependence of mediators and outcome on the unmeasured confounders, and extending the theory to nonlinear settings remains a topic for future research.}
 Third, our approach may be extended to handle causally ordered mediators using other variants of latent factor models \citep{grzebyk2004identification}. Additionally, the identification and asymptotic results in this work are derived under a fixed mediator dimension $p$. Under this setting, the unmeasured confounder $U$ cannot be recovered or consistently estimated, even if its pseudo proxy variable $L$ were known. In high-dimensional settings where $p$ grows with $n$, the theoretical framework could potentially be extended using high-dimensional factor models \citep{bai2016estimation}, and in such cases $U$ may become consistently estimable under certain conditions. For example, when $U$ is scalar, consistent estimation would require the number of confounded variables to increase with $n$, which may be a strong assumption. Investigating the theoretical properties of our approach when $p$ grows with $n$, while the number of confounded mediators remains finite, thus represents an interesting direction for future research.

	\section*{Acknowledgements}
We sincerely thank the editor, associate editor, and   reviewers for their valuable comments.
	 Wei Li  was supported by  the National Key R\&D Program of China (2022YFA1008100) and the National Natural Science Foundation of China (12471269).  The
	public computing cloud from the Renmin University
	of China was used to perform the simulation and
	data analysis. 	Wei Li is the corresponding author.

\bibliographystyle{imsart-nameyear}
\bibliography{ref.bib}

\section*{Appendix}
This Appendix provides proofs of all theoretical results presented in the main paper. It also includes additional examples, simulation studies, and  identifiability results under nonlinear outcome model settings.
\begin{appendix}

\setcounter{equation}{0} \renewcommand{\theequation}{A\arabic{equation}}

\setcounter{table}{0} \renewcommand{\thetable}{A\arabic{table}}

\setcounter{example}{0} \renewcommand{\theexample}{A\arabic{example}}

\section{Proof of Lemma \ref{lem:causal-to-stat}}\label{append:causal-to-stat}

{\bf Proof sketch:} The key idea is to express the relevant potential outcomes using iterated expectations and the latent sequential ignorability assumption, which allow conditioning on $(X,U)$ to remove dependencies involving the mediators. Substituting the linear models for $Y$ and $M$, integrating over $(X,U)$, and comparing the resulting expressions for different intervention values $(z,z')$ yield closed-form expressions for the NDE, NIE, and component-wise mediation effects. Details are provided below.

We first focus on the expected  potential outcome $E[Y\{z,M(z')\}]$ for two possibly different levels $z,z'$. Note that under Assumption \ref{ass:latent}, we have
\begin{align*}
    &E[Y\{z,M(z')\}]\\
    =&\int_m\int_x\int_u E\{Y(z,m)\mid x,u,M(z')=m\} f\{M(z')=m\mid x,u\}f(x,u)dmdxdu\\
    =&\int_m\int_x\int_u E\{Y(z,m)\mid x,u\}f(M=m\mid z',x,u)f(x,u)dmdxdu\\
    =&\int_m\int_x\int_u E(Y\mid z,m,x,u)f(M=m\mid z',x,u)f(x,u)dmdxdu,
\end{align*}
where the second equality holds because of Assumptions \ref{ass:latent}(iv) and \ref{ass:latent}(i), and the third equality holds because of Assumptions \ref{ass:latent}(ii) and \ref{ass:latent}(iii).

Under models~\eqref{eqn:modelforY}--\eqref{eqn:modelforM}, we then have
\begin{equation}\label{eqn:potentialYM}
\begin{aligned}
    &E[Y\{z,M(z')\}]\\
    =&\int_m\int_x\int_u (\beta_0 + \beta_1 z + \beta_2^{\T} m + \beta_3^{\T} x + \varphi^{\T} u)f(M=m\mid z',x,u)f(x,u)dmdxdu\\
    =&\int_x\int_u\{\beta_0 + \beta_1 z +\beta_2^{\T}E(M\mid z',x,u)+\beta_3^{\T} x + \varphi^{\T} u\}f(x,u)dxdu\\
    =&\int_x\int_u\big[\beta_0 + \beta_1 z +\beta_2^{\T}\{g(z',x)+\Gamma u\}+\beta_3^{\T} x + \varphi^{\T} u\big]f(x,u)dxdu\\
    =&\beta_0 + \beta_1 z +\beta_2^{\T}E\{g(z',X)\} + (\beta_2^{\T}\Gamma +\varphi^{\T}) E(U)+\beta_3^{\T} E(X).
\end{aligned}
\end{equation}
Based on this equation and definitions of NDE and NIE, we have
\begin{align*}
  \mathrm{NDE}&=E[ Y\{ z,M(z') \} - Y\{ z',M(z') \}]=\beta_1 (z-z'),\\
  \mathrm{NIE} &=E [ Y\{ z,M(z) \} - Y\{ z,M(z') \} ]=\beta_2^{\T} E \{ g(z,X) - g(z',X) \}.
\end{align*}

Next, we consider $E[Y\{z,M_j(z')\}]$. By the composition assumption \citep{pearl2009causality}, we have $E[Y\{z,M_j(z')\}]=E[Y\{z,M_j(z'),M_{-j}(z,M_j(z'))\}]$. In addition, by the parallel-mediator structure in Assumption \ref{ass:parallel}, we have $M_{-j}(z,m_j)=M_{-j}(z)$ and $M_j(z)\indep M_{-j}(z')\mid X,U$.
Then similar to \eqref{eqn:potentialYM}, we obtain that
\begin{align*}
    &E[Y\{z,M_j(z')\}]\\
    =&E[Y\{z,M_j(z'),M_{-j}(z,M_j(z'))\}]\\
    =&\int_m\int_x\int_u E\{Y(z,m_j,m_{-j})\mid x,u,M_j(z')=m_j, M_{-j}(z,m_j)=m_{-j}\}\\
    &~~~~~~~~~~~~\times f(M_j(z')=m_j, M_{-j}(z,m_j)=m_{-j}\mid x,u)f(x,u)dmdxdu\\
    =&\int_m\int_x\int_u E\{Y(z,m_j,m_{-j})\mid x,u\} f(M_j(z')=m_j, M_{-j}(z)=m_{-j}\mid x,u)f(x,u)dmdxdu\\
    =&\int_m\int_x\int_u E(Y\mid z,m,x,u)f(M_j=m_j\mid z',x,u)f(M_{-j}=m_{-j}\mid z,x,u)f(x,u)dmdxdu,
\end{align*}
where the third equality holds because of Assumption \ref{ass:latent}(iv), and the last equality holds because of Assumptions \ref{ass:latent}(i)--(iii) and the parallel-mediator structure.

Under models \eqref{eqn:modelforY} and \eqref{eqn:modelforM}, we then have
\begin{align*}
    &E[Y\{z,M_j(z')\}]\\
    =&\int_m\int_x\int_u (\beta_0 + \beta_1 z + \beta_2^{\T} m + \beta_3^{\T} x + \varphi^{\T} u)f(m_j\mid z',x,u)f(m_{-j}\mid z,x,u)f(x,u)dmdxdu\\
    =& \beta_0 + \beta_1 z+\int_x\int_u\big\{\beta_{2j}E(M_j\mid z',x,u)+\beta_{2,-j}^{\T}E(M_{-j}\mid z,x,u)\big\}dxdu + \beta_3^{\T} E(X) + \varphi^{\T} E(U)\\
    =&\beta_0 + \beta_1 z+\int_x\int_u\big[\beta_{2j}\{g_j(z',x)+(\Gamma u)_j\}+\beta_{2,-j}^{\T}\{g_{-j}(z,x)+(\Gamma u)_{-j} \}\big]dxdu + \beta_3^{\T} E(X) + \varphi^{\T} E(U)\\
    =&\beta_0 + \beta_1 z+\beta_{2j}E\{g_j(z',X)\}+\beta_{2,-j}^{\T}E\{g_{-j}(z,X)\}+\beta_3^{\T} E(X)+(\beta_2^{\T}\Gamma +\varphi^{\T}) E(U).
\end{align*}
Therefore, by the definition of $\mathrm{NIE}_j$, we have
\begin{align*}
    \mathrm{NIE}_j = E[ Y\{ z,M_j(z) \} - Y\{ z,M_j(z') \} ]=\beta_{2j} E \{ g_j(z,X) - g_j(z',X) \}.
\end{align*}

\section{Proof of Theorem~\ref{THM:IDENTIFICATION_1}} \label{proof_thm1}

{\textbf{Proof sketch:} by expressing the outcome in terms of the pseudo proxy variable $L$ and a residual $\psi$, we show that $\psi$ is uncorrelated with $Z$, $M$, $X$, and $L$. Using standard linear algebra and the full-rank condition, we then establish that the regression coefficients $(\beta_0, \beta_1, \beta_2, \beta_3)$ are identifiable, and $\Gamma$ and $\varphi$ are identifiable up to rotation. Details are provided below.}

Denote $ \Delta =(\Sigma_{\varepsilon} + \Gamma \Gamma^\T)^{-1} \Gamma, \check{M} = M - g(Z,X) = \Gamma U +\varepsilon, L = \Delta^\T \check{M}$ and we have
\begin{align*}     
	Y& = \beta_0 +\beta_1 Z +\beta_2^\T M + \beta_3^\T X + \varphi^\T L +\eta +\varphi^\T U - \varphi^\T L \\
	& =\beta_0 + \beta_1 Z + \beta_2^\T M + \beta_3^\T X + \varphi^\T L + \psi,
\end{align*}
where $\psi = \eta +\varphi^\T U -b^\T(\Gamma U +\varepsilon)$ and $b=\Delta \varphi$.

Based on the full rank condition of Theorem 1, we firstly verify that $\psi$ is uncorrelated with $Z,X,M
$ and $L$ for the identifiability of $\beta_2$. We have

\begin{itemize}
	\item[i)] 
	\begin{align*}
		\mathrm{cov} (\psi,\check{M}) & = \mathrm{cov} \big\{ \eta +\varphi^\T U -b^\T(\Gamma U +\varepsilon), \Gamma U +\varepsilon \big\} \\
		& =  \mathrm{cov} ( \varphi^\T U, \Gamma U ) - b^\T (\Gamma \Gamma^\T +\Sigma_{\varepsilon} )\\
		& = \varphi^\T \Gamma^\T - \varphi^\T \Delta^\T ( \Gamma \Gamma^\T +\Sigma_{\varepsilon} )\\
		& = \varphi^\T \Gamma^\T - \varphi^\T \Gamma^\T\\
		&=0.
	\end{align*}
	
	\item[ii)]
	\begin{align*}
		\mathrm{cov} (\psi,X)& = \mathrm{cov} \big\{ \eta +\varphi^\T U -b^\T ( \Gamma U +\varepsilon ), X \big\}\\
		&=0.
	\end{align*}
	
	\item[iii)]
	\begin{align*}
		\mathrm{cov} (\psi,Z) &= \mathrm{cov} \big\{ \eta +\varphi U -b^\T, (\Gamma U +\varepsilon ), Z \big\}\\
		&=0.
	\end{align*}
	
	\item[iv)]
	\begin{align*}
		\mathrm{cov} (\psi,M) & = \mathrm{cov} \big\{ \eta +\varphi^\T U -b^\T ( \Gamma U +\varepsilon ),\check{M} +  g(Z,X) \big\}\\
		&=0.
	\end{align*}
\end{itemize}
We immediately know that $\mathrm{cov} (\psi, L) = \mathrm{cov} (\psi, \Delta^\T \check{M}) = 0$. So we have proved that $Z,M,X$ and $L$ are uncorrelated with $\psi$. Then $\beta = (\beta_0, \beta_1, \beta_2^\T, \beta_3^\T, \varphi^\T)^\T$ can be represented as
\begin{align*}
	\beta = \big[ E \{ ( 1,Z,M^\T,X^\T,L^\T )^\T ( 1,Z,M^\T,X^\T,L^\T ) \} \big]^{-1} E \{ ( 1,Z,M^\T,X^\T,L^\T)^\T Y \}.
\end{align*}
Let
\begin{align*}
	E\{ ( 1,Z,M^\T,X^\T,L^\T )^\T ( 1,Z,M^\T,X^\T,L^\T ) \}
	= 
	\begin{pmatrix}
		A & B  \\ 
		B^\T & C  
	\end{pmatrix},
\end{align*}
and 
\begin{align*}
	E\{ ( 1,Z,M^\T,X^\T,L^\T )^\T Y \} 
	= 
	\begin{pmatrix}
		D \\
		F 
	\end{pmatrix},
\end{align*}
where 
\begin{align*}
	&A = E\{ ( 1,Z,M^\T,X^\T )^\T ( 1,Z,M^\T,X^\T ) \}, ~ B = E\{ ( 1,Z,M^\T,X^\T )^\T L^\T \}, \\ 
	&C = E( L L^\T ), ~ D = E\{ ( 1,Z,M^\T,X^\T )^\T Y \}, ~ F = E( L Y ).
\end{align*}
It is easy to observe that
$$
\begin{pmatrix}
	A & B  \\ 
	B^\T & C  
\end{pmatrix}^{-1} = 
\begin{pmatrix}
	(A-B C^{-1} B^\T )^{-1} & - A^{-1} B (C - B^\T A^{-1} B)^{-1} \\
	-C^{-1} B^\T (A - B C^{-1} B^\T)^{-1} & (C - B^\T A^{-1} B)^{-1}
\end{pmatrix}.
$$
Now we know that
\begin{align*}(\beta_0, \beta_1, \beta_2^\T, \beta_3^\T) 
	&= 
	\begin{pmatrix}
		(A - B C^{-1} B^\T )^{-1} & - A^{-1} B (C - B^\T A^{-1} B)^{-1}
	\end{pmatrix} 
	\begin{pmatrix}
		D \\
		F
	\end{pmatrix} \\
	&= (A-B C^{-1} B^\T )^{-1} D  - A^{-1} B (C - B^\T A^{-1} B)^{-1} F.
\end{align*}
Firstly, we know $\Gamma$ is identifiable up to some rotation, implying that $\Gamma$ may be replaced by $\Gamma S $, where S is an orthogonal matrix. So the predictor $L = \Delta^\T \check{M}$ may be replaced by $S^\T L$, and $B,C$ $F$ will be respectively replaced by $BS, S^\T C S$ and $ S^\T F$. Based on this, we immediately know $B C^{-1} B^\T$ and $B(C- B^\T A^{-1} B )^{-1} F$ are both identifiable, which implies that $\beta_0, \beta_1, \beta_2^\T, \beta_3^\T$ are all identifiable. Thus, if we replace $L$ with $S^\T L$, where $S$ is the rotation matrix, our identification results are still valid. Similarly, we can immediately conclude that $\varphi$ is also identifiable up to some rotation. This completes the proof of Theorem 1.\\

\section{Proof of Theorem~\ref{THM:NORMAL}}  \label{proof_thm2}

{\textbf{Proof sketch:} by expanding $\hat{L}_i$ via a Taylor series and applying empirical process theory, we show that the contributions from estimating $L_i$ vanish asymptotically, and the penalization term is negligible for $\lambda_n = o(\sqrt{n})$. Using the Argmin Theorem, we then establish that the lasso estimator $\hat{\xi}_{\mathrm{la}}$ is asymptotically normal with limiting covariance $\Sigma_{\mathrm{la}}$. Details are provided below.
}

Let
$$
Z_n(\phi) =\dfrac{1}{n} \sum_{i=1}^n \big( Y_i - \phi_0 -\phi_1 Z_i - \phi_2^\T M_{i \cdot } - \phi_3^\T X_i - \phi_4^\T \hat{L}_i \big)^2 + \dfrac{\lambda_n}{n} \lVert \phi_2 \rVert_1,
$$
where $\hat{L}_i = \hat{\Delta}^\T \check{M}_i ( \hat{\gamma} ), \hat{\Delta} =\big( \hat{\Sigma}_{\varepsilon} + \hat{\Gamma} \hat{\Gamma}^\T \big)^{-1} \hat{\Gamma}$ and $\check{M}_i ( \hat{\gamma} ) = M_{i \cdot} -  g(Z_i, X_i; \hat{\gamma})$. From~\citet{hansen1982large}, we have
$$ 
\sqrt{n} \big( \hat{\nu} - \nu_0 \big) \overset{d}{\to}  N(0,B),
$$
where 
$$
B = \bigg[ \dfrac{ \partial E\{ Q(S; \nu_0) \} } { \partial \nu } \bigg]^{-1} \mathrm{var}\{ Q(S;\nu_0) \} \bigg[ \dfrac{ \partial E\{ Q(S;\nu_0) \} } { \partial \nu } \bigg]^{-\T}.
$$
Thus, the estimators $\hat{\nu}$ are asymptotically jointly normal. Now suppose $\lambda_n = o(\sqrt{n})$, we have
\begin{align*}
	n Z_n (\phi) & =  \sum_{i=1}^n \big( Y_i - \phi_0 - \phi_1 Z_i - \phi_3^\T X_i - \phi_2^\T M_{i \cdot} - \phi_4^\T \hat{L}_i \big)^2 + \lambda_n \lVert \phi_2 \rVert_1   \\
	& =  \sum_{i=1}^n   \bigg\{ \psi_i - \dfrac{1}{\sqrt{n}}   \sqrt{n} (\phi - \xi)^\T  R_i + \dfrac{1}{\sqrt{n}} \sqrt{n} \phi_4^\T \big( L_i - \hat{L}_i \big) \bigg\} ^2   + \lambda_n \lVert \phi_2 \rVert_1 ,     
\end{align*}
where $R_i = (1, Z_i, M_{i \cdot}^\T, X_i^\T, L_i^\T)^\T$.

Hence, we construct the following function whose unique minimizer is $\sqrt{n} \big ( \hat{\xi}_{\mathrm{la}}- \xi \big)$:
$$
V_n(u) = \sum_{i=1}^n \bigg[ \big\{ \psi_i - \dfrac{R_i^\T u}{ \sqrt{n} } - \big( \hat{L}_i - L_i \big)^\T \big( \dfrac{u_4}{\sqrt{n}} +\varphi \big) \big\} ^2 - (\psi_i)^2   \bigg]  + \lambda_n \bigg\{ \lVert \beta_2 + \dfrac{u_2} {\sqrt{n} } \rVert_1 - \lVert \beta_2 \rVert_1  \bigg\}.
$$
Now it is easy to see that $\lambda_n \big( \lVert \beta_2 + u_2 / \sqrt{n}  \rVert_1 - \lVert \beta_2 \rVert_1  \big) \overset{n\to\infty}{\longrightarrow} 0$. Also, we have:
\begin{align*}
	& \sum_{i=1}^n \bigg[ \big\{ \psi_i - \dfrac{R_i^\T u}{ \sqrt{n} } - \big( \hat{L}_i - L_i \big)^\T \big( \dfrac{u_4}{\sqrt{n}} +\varphi \big) \big\} ^2 - (\psi_i)^2   \bigg]  \\
	= & u^\T \bigg( \dfrac{1}{n} \sum_{i=1}^n R_i R_i^\T \bigg) u + u_4^\T \bigg\{ \dfrac{1}{n} \sum_{i=1}^n \big( \hat{L}_i - L_i \big) \big( \hat{L}_i - L_i \big)^\T  \bigg\}  u_4 - 2 u^\T \bigg( \dfrac{1}{\sqrt{n}} \sum_{i=1}^n R_i \psi_i \bigg) - 2u_4^\T \\
	&\cdot \bigg\{ \dfrac{1}{\sqrt{n}} \sum_{i=1}^n \big( \hat{L}_i - L_i \big) \psi_i  \bigg\} + 2u^\T \bigg\{ \dfrac{1}{n} \sum_{i=1}^n R_i \big( \hat{L}_i - L_i \big)^\T \bigg\} u_4 + 2u_4^\T \bigg\{ \dfrac{1}{\sqrt{n}} \sum_{i=1}^n \big( \hat{L}_i - L_i \big)  \big( \hat{L}_i - L_i \big)^\T \bigg\} \varphi \\
	& + 2u^\T \bigg\{ \dfrac{1}{\sqrt{n}} \sum_{i=1}^n R_i \big( \hat{L}_i - L_i \big)^\T \bigg\} \varphi +  \mathrm{const.} \\
	= & \sum_{i=1}^7 T_i + \mathrm{const.}
\end{align*}
Now we ignore the constant in $V_n(u)$ and firstly by the empirical process theory in \citet{pollard1989asymptotics} and \citet{andrews1994empirical} we have
$$
T_2,T_5 \overset{P}{\longrightarrow} 0, ~ T_1 \overset{P}{\longrightarrow} u^\T C u.
$$
We assume $L\in \mathbb{R}^t$ and check the Taylor expansion of every item of $\hat{L}$, the $j$-th item $\hat{L}(j)$. Then we can use such expansion to prove that $T_4, T_6 \to 0$, now because
\begin{align*}
	\hat{L}(j) = L(j;\nu_0) + \dfrac{\partial L(j;\nu_0)}{\partial \nu^\T} \big( \hat{\nu} - \nu_0 \big)  + \dfrac{1}{2n} \bigg\{ \sqrt{n} \big( \hat{\nu} -\nu_0 \big)^\T  \dfrac{\partial^2 L\big( j;\Tilde{\nu} \big) }{\partial \nu \partial \nu^\T}   \sqrt{n} \big( \hat{\nu} -\nu_0 \big) \bigg\}.
\end{align*}
We immediately have (sometimes we use $L(j)$ instead of $L(j;\nu_0)$ for short)
\begin{align*}
	& \dfrac{1}{\sqrt{n}} \sum_{i=1}^n \big\{ \hat{L}_i(j) - L_i(j) \big\} \psi_i  \\
	=  & \dfrac{1}{\sqrt{n}} \sum_{i=1}^n \bigg[  \dfrac{\partial L_i(j;\nu_0)}{\partial \nu^\T}  \big( \hat{\nu} - \nu_0 \big)  + \dfrac{1}{2n} \big\{ \sqrt{n} (\hat{\nu} -\nu_0 )^\T  \dfrac{\partial^2 L_i \big( j;\Tilde{\nu}_i \big)}{\partial \nu \partial \nu^\T}  \sqrt{n} \big( \hat{\nu} -\nu_0 \big) \big\}  \bigg] \psi_i .
\end{align*}
Firstly we can show
$$
\dfrac{1}{n} \sum_{i=1}^n  \dfrac{ \partial^2 L_i \big( j;\Tilde{\nu}_i \big) } { \partial \nu \partial \nu^\T } \psi_i  = E \bigg\{  \dfrac{\partial^2 L(j;\nu_0)} { \partial \nu \partial \nu^\T }  \psi \bigg\}   + o_p(1) =  \dfrac{\partial^2}{\partial \nu \partial \nu^\T}  E\{ L(j;\nu_0)  \psi  \}    + o_p(1) = o_p(1) ,
$$
which can be derived from~\citet{serfling2009approximation}. Then we know
\begin{align*}
	& \dfrac{1}{\sqrt{n}} \sum_{i=1}^n \big\{ \hat{L}_i(j) - L_i(j;\nu_0) \big\} \psi_i \\
	= & \dfrac{1}{n} \sum_{i=1}^n \bigg\{ \dfrac{\partial L_i(j;\nu_0)}{\partial \nu}  \psi_i \bigg\}  \sqrt{n} \big( \hat{\nu} - \nu_0 \big)  +  \sqrt{n} \big( \hat{\nu} -\nu_0 \big)^\T  \dfrac{o_p(1)}{\sqrt{n}}   \sqrt{n} \big( \hat{\nu} -\nu_0 \big),
\end{align*}
which implies
\begin{align*}
	\dfrac{1}{\sqrt{n}} \sum_{i=1}^n \big\{ \hat{L}_i(j) - L_i(j;\nu_0) \big\} \psi_i \overset{d}{\to} E \bigg\{ \dfrac{\partial L_i(j;\nu_0) }{\partial \nu^\T}  \psi \bigg\} \cdot O_p(1) = \dfrac{\partial}{\partial \nu^\T} E\{ L(j;\nu_0)  \psi \} \cdot O_p(1) = 0 , 
\end{align*}
Thus, the Slutsky's Theorem immediately implies
\begin{align*}
	\dfrac{1}{n} \sum_{i=1}^n  \dfrac{\partial L_i (j;\nu_0)}{\partial \nu^\T} \psi_i \cdot \sqrt{n} \big( \hat{\nu} - \nu_0 \big)  & \overset{d}{\to} 0 \cdot N (0,B) = 0 , \\
	\sqrt{n} \big( \hat{\nu} -\nu_0 \big)^\T \cdot \dfrac{o_p(1)}{\sqrt{n}} \cdot  \sqrt{n} \big( \hat{\nu} -\nu_0 \big)^\T & \overset{d}{\to} N (0, B)^\T \cdot 0 \cdot N  (0,B) = 0 .
\end{align*}
Combined from all above, we have
$$
T_4 = - 2u_4^\T \bigg\{ \dfrac{1}{\sqrt{n}} \big( \hat{L}_i - L_i \big) \psi_i  \bigg\} = -2 \sum_{j=1}^t u_4(j) \dfrac{1}{\sqrt{n}} \sum_{i=1}^n  \big\{ \hat{L}_i(j) - L_i(j;\nu_0) \big\} \psi_i \overset{d}{\to} 0 .
$$
Similarly we can obtain the Taylor expansion of the $(h,k)$th item of $(\hat{L} - L) (\hat{L} - L)^\T$:
\begin{align*}
	& \big\{ \hat{L}(h) - L(h;\nu_0) \big\} \big\{ \hat{L}(k) - L(k;\nu_0) \big\} \\
	= & \big\{ \hat{L}(h) - L(h;\nu_0) \big\} \big\{ \hat{L}(k) - L(k;\nu_0) \big\}\rvert_{\nu_0}  + \bigg[ \dfrac{\partial L(h;\nu_0)}{\partial \nu^\T} \big\{ \hat{L}(k) - L(k;\nu_0) \big\} + \dfrac{\partial L(k;\nu_0)}{\partial \nu^\T} \\
	& \cdot \big\{ \hat{L}(h) - L(h;\nu_0) \big\} \bigg]\rvert_{\nu_0}  (\hat{\nu} - \nu_0 ) + \dfrac{1}{2} \big( \hat{\nu} - \nu_0 \big)^\T   \dfrac{\partial^2 \big\{ \hat{L}(h) - L(h;\nu_0) \big\} 
		\big\{ \hat{L}(k) - L(k;\nu_0) \big\}}{\partial \nu \partial \nu^\T} \rvert_{\Tilde{\nu}} \\
	& \cdot (\hat{\nu} - \nu_0).
\end{align*}
It is obvious that the first and the second item is zero. Now we need to check the second derivative:
\begin{align*}
	& \dfrac{\partial^2 \big\{ \hat{L}(h) - L(h;\nu_0) \big\} \big\{ \hat{L}(k) - L(k;\nu_0) \big\} }{\partial \nu \nu^\T}\rvert_{\Tilde{\nu}} \\
	=  & \bigg[   \dfrac{\partial \hat{L}(h) }{\partial \nu}  \dfrac{\partial \hat{L}(k) }{\partial \nu^\T} + \dfrac{\partial^2  \hat{L} (h)}{\partial \nu \partial \nu^\T}  \big\{ \hat{L}(k) - L(k;\nu_0) \big\}  + \dfrac{\partial \hat{L}(k) }{\partial \nu}  \dfrac{\partial \hat{L}(h) }{\partial \nu^\T} + \dfrac{\partial^2  \hat{L} (k)}{\partial \nu \partial \nu^\T}   \big\{ \hat{L}(h) - L(h;\nu_0) \big\}
	\bigg]\rvert_{\Tilde{\nu}}.
\end{align*}
Then we must know:
\begin{align*}
	&\dfrac{1}{\sqrt{n}} \sum_{i=1}^n \big\{ \hat{L}_i(h) - L_i(h;\nu_0) \big\}  \big\{ \hat{L}_i(k) - L_i(k;\nu_0) \big\}\\
	=  & \sqrt{n} \big( \hat{\nu} - \nu_0 \big)^\T   \dfrac{1}{n\sqrt{n}} \sum_{i=1}^n  \bigg[   \dfrac{\partial \hat{L}_i(h) }{\partial \nu}  \dfrac{\partial \hat{L}_i(k) }{\partial \nu^\T} + \dfrac{\partial^2  \hat{L}_i (h)}{\partial \nu \partial \nu^\T}  \big\{ \hat{L}_i(k) - L_i(k;\nu_0) \big\}  + \dfrac{\partial \hat{L}_i(k) }{\partial \nu}  \dfrac{\partial \hat{L}_i(h) }{\partial \nu^\T} \\
	+ & \dfrac{\partial^2  \hat{L}_i (k)}{\partial \nu \partial \nu^\T}  \big\{ \hat{L}_i(h) - L_i(h;\nu_0) \big\}
	\bigg]\rvert_{\Tilde{\nu}_i}   
	\sqrt{n}(\hat{\nu} - \nu_0) .
\end{align*}
Now we also know:
\begin{align*}
	&\dfrac{1}{n\sqrt{n}} \sum_{i=1}^n \bigg[   \dfrac{\partial \hat{L}_i(h) }{\partial \nu}  \dfrac{\partial \hat{L}_i(k) }{\partial \nu^\T}  + \dfrac{\partial^2 \hat{L}_i(h) }{\partial \nu \partial \nu^\T} \big\{ \hat{L}_i(k)  - L_i(k;\nu_0) \big\} \bigg]\rvert_{\Tilde{\nu}_i} \\
	= &\dfrac{1}{\sqrt{n}} E \bigg[   \dfrac{\partial L(h;\nu_0)}{\partial \nu}  \dfrac{\partial L(k;\nu_0)}{\partial \nu^\T} + \dfrac{\partial^2 \hat{L}(h) }{\partial \nu \partial \nu^\T}  \big\{ \hat{L}(k) - L(k;\nu_0) \big\} \bigg]\rvert_{\nu_0} + o_p(1) \\
	= &\dfrac{1}{\sqrt{n}} E\bigg\{ \dfrac{\partial L(h;\nu_0)}{\partial \nu}  \dfrac{\partial L(k;\nu_0)}{\partial \nu^\T}  \bigg\}  + o_p(1) \\
	=& o_p(1) ,
\end{align*}
which is also derived from~\citet{serfling2009approximation}, and the above relation also holds when we interchange $h$ and $k$. Now we have:
$$ 
\dfrac{1}{\sqrt{n}} \sum_{i=1}^n  \big\{ \hat{L}_i(h) - L_i(h;\nu_0) \big\} \big\{ \hat{L}_i(k) - L_i(k;\nu_0) \big\}
\overset{d}{\to} N(0,B)^\T \cdot 0 \cdot N (0, B) = 0 ,
$$
which implies
\begin{align*}
	T_6 &= 2u_4^\T \bigg\{ \dfrac{1}{\sqrt{n}} \sum_{i=1}^n  \big( \hat{L}_i - L_i \big) \big( \hat{L}_i - L_i \big)^\T \bigg\} \varphi \\
	&= 2 \sum_{h,k =1}^t u_4(h)\varphi(k) \dfrac{1}{\sqrt{n}} \sum_{i=1}^n \big\{ \hat{L}_i(h) - L_i(h;\nu_0) \big\} \big\{ \hat{L}_i(k) - L_i(k;\nu_0) \big\} \\
	& \overset{d}{\to} 0 .
\end{align*}
Finally, we will check the limiting distribution of $T_3 + T_7$, firstly we have
\begin{align*}
	& \dfrac{1}{\sqrt{n}} \sum_{i=1}^n R_i \big\{ \hat{L}_i(j) - L_i(j;\nu_0) \big\} \\
	= & \dfrac{1}{\sqrt{n}} \sum_{i=1}^n  R_i  \bigg[  \dfrac{\partial L_i(j;\nu_0)}{\partial \nu^\T} \big( \hat{\nu} - \nu_0 \big)  + \dfrac{1}{2n} \big\{ \sqrt{n} \big( \hat{\nu} -\nu_0 \big)^\T  \dfrac{\partial^2 \hat{L}_i(j)}{\partial \nu \partial \nu^\T}\rvert_{\Tilde{\nu}_i}   \sqrt{n} \big( \hat{\nu} -\nu_0 \big)   \big\} \bigg] .
\end{align*}
Furthermore, we can show from~\citet{serfling2009approximation}
$$
\dfrac{1}{n} \sum_{i=1}^n \dfrac{\partial^2 \hat{L}_i(j) }{\partial \nu \nu^\T}\rvert_{\Tilde{\nu}_i} 
= E \bigg\{ \dfrac{\partial^2 L(j;\nu_0)}{\partial \nu \nu^\T} \bigg\} + o_p(1) = o_p(1).
$$
Because
$$
\dfrac{1}{n} \sum_{i=1}^n R_i   \dfrac{\partial L_i(j;\nu_0)}{\partial \nu^\T}  \overset{d}{\to}  E \bigg\{ R \dfrac{\partial L(j;\nu_0)}{\partial \nu^\T} \bigg\},
$$
then we have
\begin{align*}
	& \dfrac{1}{\sqrt{n}} \sum_{i=1}^n R_i  \dfrac{\partial L_i(j;\nu_0)}{\partial \nu}  \big( \hat{\nu} - \nu_0 \big) \\
	= & \bigg\{  \dfrac{1}{n} \sum_{i=1}^n R_i \dfrac{\partial L_i(j;\nu_0)}{\partial \nu^\T} \bigg\}  \sqrt{n} \big( \hat{\nu} - \nu_0 \big) \\
	= & E \bigg\{  R \dfrac{\partial L(j;\nu_0)}{\partial \nu^\T} \bigg\}  \sqrt{n} \big( \hat{\nu} - \nu_0 \big) + o_p(1) ,
\end{align*}
and
\begin{align*}
	\dfrac{1}{\sqrt{n}} \sum_{i=1}^n  \dfrac{1}{2n} \bigg\{ \sqrt{n} \big( \hat{\nu} -\nu_0 \big)^\T  \dfrac{\partial^2 \hat{L}_i(j)}{\partial \nu \partial \nu^\T}\rvert_{\Tilde{\nu}_i}   \sqrt{n} \big( \hat{\nu} -\nu_0 \big)   \bigg\}   \overset{d}{\to}  N(0,B)^\T \cdot 0 \cdot N(0,B) = 0.
\end{align*}
So we know
\begin{align*}
	T_3 + T_7  
	= & 2u^\T \bigg\{ \dfrac{1}{\sqrt{n}} \sum_{i=1}^n R_i \big( \hat{L}_i - L_i \big)^\T\varphi - \dfrac{1}{\sqrt{n}} \sum_{i=1}^n R_i \psi_i \bigg\} \\
	= & 2u^\T \bigg[ \sum_{j=1}^t \dfrac{1}{\sqrt{n}} \sum_{i=1}^n R_i \big\{ \hat{L}_i(j) - L_i(j;\nu_0) \big\} \varphi(j) - \dfrac{1}{\sqrt{n}} \sum_{i=1}^n R_i \psi_i \bigg]    \\
	= & 2u^\T \bigg[ \sum_{j=1}^t \varphi(j)  E \bigg\{  R \dfrac{\partial L(j;\nu_0)}{\partial \nu^\T} \bigg\}  \sqrt{n} \big( \hat{\nu} - \nu_0 \big) + o_p(1) - \dfrac{1}{\sqrt{n}} \sum_{i=1}^n R_i \psi_i   \bigg]\\
	= & 2u^\T \bigg( -   E \bigg[  R \dfrac{\partial \{ \varphi^\T L(\nu_0) \}}{\partial \nu} \bigg] \bigg[ \dfrac{\partial E\{ Q(S;\nu_0) \} }{\partial \nu} \bigg]^{-1} \dfrac{1}{\sqrt{n}} \sum_{i=1}^n Q(S_i;\nu_0) - \dfrac{1}{\sqrt{n}} \sum_{i=1}^n R_i \psi_i  + o_p(1) \bigg)\\
	= & -2u^\T \bigg\{ \dfrac{1}{\sqrt{n}} \sum_{i=1}^n \bigg(   E \bigg[  R \dfrac{\partial \{ \varphi^\T L(\nu_0) \}}{\partial \nu^\T} \bigg]  \bigg( \dfrac{\partial E\{ Q(S;\nu_0) \}}{\partial \nu} \bigg)^{-1}  Q(S_i;\nu_0) + R_i \psi_i \bigg) + o_p(1)   \bigg\}\\
	\to & -2u^\T K ,
\end{align*}
where $K\sim N(0,\Sigma_K), \Sigma_K = E( K K^\T )$ and
$$
K =   E \bigg[  R \dfrac{\partial \{ \varphi^\T L(\nu_0)\}}{\partial \nu^\T} \bigg]  \bigg( \dfrac{\partial E\{ Q(S;\nu_0) \}}{\partial \nu} \bigg)^{-1}  Q(S;\nu_0) + R \psi  .
$$
So we have
$$
V_n(u) \overset{d}{\longrightarrow} u^\T C u - 2u^\T K .
$$
Now it is obvious to see
$$ 
\sqrt{n} \big( \hat{\xi}_{\mathrm{la}}  - \xi \big) = \mathop{\arg\min} V_n(u) \overset{d}{\longrightarrow} \mathop{\arg\min} V(u) = W , $$
where $W= C^{-1} K\sim N(0,\Sigma_{\mathrm{la}}) $ with $\Sigma_{\mathrm{la}} = C^{-1} E( K K^\T ) C^{-1} $. This is derived by the Argmin Theorem~\citep{geyer1996asymptotics}. Thus, we have proved that the estimators are asymptotically normal.

\section{Proof of Theorem~\ref{THM:AD_NORMAL}}  \label{proof_thm3}

{\textbf{Proof sketch:} 
the proof proceeds by reparametrizing the adaptive lasso objective in terms of $u = \sqrt{n} (\hat{\xi}_{\mathrm{ad}} - \xi)$ and analyzing its limiting behavior. For nonzero coefficients, the adaptive penalty vanishes asymptotically, while for zero coefficients, it diverges unless the estimate is exactly zero, ensuring sparsity. Applying convexity and the argmin theorem, the limiting objective has a unique minimizer, yielding asymptotic normality for the active coefficients and exact zeros for inactive ones. Finally, using the KKT conditions, it is shown that the probability of incorrectly selecting zero coefficients vanishes, establishing both asymptotic normality and selection consistency of the adaptive lasso estimator. Details are provided below.
}

Similarly as the proof of Theorem \ref{THM:NORMAL}, let
$$
n D_n(\phi) = \sum_{i=1}^n \big( Y_i - \phi_0 - \phi_1 Z_i - \phi_2^\T M_{i \cdot} - \phi_3^\T X_i - \phi_4^\T \hat{L}_i   \big)^2 + \lambda_n \sum_{r=1}^p \hat{w}_r \lvert \phi_{2,r} \rvert,
$$
where $\hat{L}_i = \hat{\Delta} \check{M}_i(\hat{\gamma})$ and $\check{M}_i(\hat{\gamma}) = M_{i \cdot} - g(Z_i,X_i;\hat{\gamma})$. Similar to the proof of Theorem \ref{THM:NORMAL}, we construct a function whose unique minimizer is $\sqrt{n} \big( \hat{\xi}_{\mathrm{ad}} - \xi \big)$:
$$
B_n(u) = \sum_{i=1}^n \bigg[ \big\{ \psi_i - \dfrac{R_i^\T u}{ \sqrt{n} } - \big( \hat{L}_i - L_i \big)^\T \big( \dfrac{u_4}{\sqrt{n}} +\varphi \big) \big\} ^2 - (\psi_i)^2   \bigg]  + \lambda_n \sum_{r=1}^p  \hat{w}_r \bigg( \lvert \beta_{2,r} + \dfrac{u_{2,r}} {\sqrt{n} } \rvert - \lvert \beta_{2,r} \rvert \bigg),
$$
where $\hat{w}=(\hat{w}_1,\cdots, \hat{w}_p)^\T = \lvert  \hat{\beta}_{\mathrm{in},2} \rvert^{-\delta}$ for some $\sqrt{n}$-consistent estimator $\hat{\beta}_{\mathrm{in},2}$ and some tuning parameter $\delta>0$. Here the first term is the same as that in the proof of Theorem \ref{THM:NORMAL} and we immediately know
$$
\sum_{i=1}^n \bigg[ \big\{ \psi_i - \dfrac{R_i^\T u}{ \sqrt{n} } - \big( \hat{L}_i - L_i \big)^\T \big( \dfrac{u_4}{\sqrt{n}} +\varphi \big) \big\} ^2 - (\psi_i)^2   \bigg]   \overset{d}{\longrightarrow} u^\T C u - 2u^\T K.
$$
Now we check the second term:
\begin{itemize}
	\item[1)] $\beta_{2,r} \neq 0$: 
	\begin{align*}
		\lambda_n  \hat{w}_r \bigg( \lvert \beta_{2,r} + \dfrac{u_{2,r}} {\sqrt{n} } \rvert - \lvert \beta_{2,r} \rvert \bigg) 
		&= \dfrac{\lambda_n}{\sqrt{n}} \cdot \hat{w}_r \cdot \sqrt{n} \bigg( \lvert \beta_{2,r} + \dfrac{u_{2,r}} {\sqrt{n} } \rvert - \lvert \beta_{2,r} \rvert \bigg) \\ 
		&\overset{d}{\longrightarrow} 0 \cdot \lvert \beta_{2,r} \rvert^{-\delta} \cdot u_{2,r} \mathrm{sgn}( \beta_{2,r} ) \\
		&= 0,
	\end{align*}
	where $\mathrm{sgn}(\cdot)$ is the sign function.
	\item[2)] $\beta_{2,r} = 0$: 
	\begin{align*}
		\lambda_n  \hat{w}_r \bigg( \lvert \beta_{2,r} + \dfrac{u_{2,r}} {\sqrt{n} } \rvert - \lvert \beta_{2,r} \rvert \bigg) 
		&= \dfrac{\lambda_n}{\sqrt{n}} \cdot \hat{w}_r \cdot \lvert u_{2,r}  \rvert  \\
		&= \dfrac{\lambda_n}{\sqrt{n}} \cdot \lvert \hat{\beta}_{2,r}  \rvert^{-\delta} \cdot \lvert u_{2,r}  \rvert \\
		&= \lambda_n n^{\dfrac{\delta -1}{2}} \lvert \sqrt{n} \hat{\beta}_{2,r}  \rvert^{-\delta}  \cdot \lvert u_{2,r}  \rvert 
	\end{align*}
	where $\lambda_n n^{(\delta-1)/2} \to \infty$ and $\sqrt{n} \hat{\beta}_{2,r}=O_p(1)$. Thus, the above term will equal 0 when $u_{2,r}=0$ while it will tend to $\infty$ if $u_{2,r}\neq 0$.
\end{itemize}
The set $\mathcal{A} = \{ j: \beta_{2j}\neq 0 \}$ and $\hat{\mathcal{A}}_n$ respectively represent the index union from the nonzero coefficients of parameter $\beta_2$ and its corresponding adaptive lasso estimator. We use $\Tilde{\mathcal{A}}$ to denote the index set union from index set of $\beta_0,\beta_1,\beta_{2,\mathcal{A}},\beta_3,\varphi$.  Now we immediately know
\begin{itemize}
	\item[1)]  $u_{2,j}=0$ for all $j\not\in \mathcal{A}$:
	$$
	\lambda_n \sum_{r=1}^p  \hat{w}_r \bigg( \lvert \beta_{2,r} + \dfrac{u_{2,r}} {\sqrt{n} } \rvert - \lvert \beta_{2,r} \rvert \bigg) \overset{d}{\longrightarrow} 0,
	$$
	\item[2)] Else:
	$$
	\lambda_n \sum_{r=1}^p  \hat{w}_r \bigg( \lvert \beta_{2,r} + \dfrac{u_{2,r}} {\sqrt{n} } \rvert - \lvert \beta_{2,r} \rvert \bigg) \longrightarrow \infty,
	$$
\end{itemize}
which implies that 
$$
B_n(u) \overset{d}{\longrightarrow} B(u) = 
\begin{cases}
	u_{\Tilde{\mathcal{A}}}^\T \Tilde{C} u_{\Tilde{\mathcal{A}}} - 2u_{\Tilde{\mathcal{A}}}^\T K_{\Tilde{\mathcal{A}}}  \quad  &\mathrm{if} ~ u_{2,j}=0, \forall j\not\in \mathcal{A} \\
	\infty    &\mathrm{otherwise.}
\end{cases},
$$
where $\Tilde{C} = C_{\Tilde{\mathcal{A}},\Tilde{\mathcal{A}}}$. Since $B_n(u)$ and $B(u)$ are both convex, from the results from~\citet{geyer1996asymptotics} and \citet{fu2000asymptotics}, the corresponding unique minimizer must satisfy
$$
\hat{u}_{\Tilde{\mathcal{A}}} \overset{d}{\longrightarrow} \Tilde{C}^{-1} K_{\Tilde{\mathcal{A}}}  , ~~ \hat{u}_{\Tilde{\mathcal{A}}^c} \overset{d}{\longrightarrow} 0,
$$
where $\hat{u} = ( \hat{u}_{\Tilde{\mathcal{A}}}, \hat{u}_{\Tilde{\mathcal{A}}^c})^\T$ and $u = (\Tilde{C}^{-1} K_{\Tilde{\mathcal{A}}},0)^\T$ are respectively the minimizer of $B_n(u)$ and $B(u)$. This implies
$$
\sqrt{n} \big( \hat{\xi}_{\mathrm{ad},\Tilde{\mathcal{A}}} -\xi_{\Tilde{\mathcal{A}}}  \big) \overset{d}{\longrightarrow} \Tilde{C}^{-1} K_{\Tilde{\mathcal{A}}}, ~~ \sqrt{n} \big( \hat{\xi}_{\mathrm{ad},\Tilde{\mathcal{A}}^c} -\xi_{\Tilde{\mathcal{A}}^c}  \big) \overset{d}{\longrightarrow} 0.
$$
The asymptotic normality result immediately implies that $\forall j\in \mathcal{A}, P(j\in \hat{\mathcal{A}}_n)\to 1$.
Now we prove the selection consistency by showing that
$$
\forall j\not\in \Tilde{\mathcal{A}}, P(j\in \mathcal{A}_n) \longrightarrow 0.
$$
Firstly, the KKT optimality condition implies
$$
2 \sum_{i=1}^n \hat{R}_{i,j} \big(  Y_i - \hat{\xi}^\T \hat{R}_i \big) = \lambda_n \hat{w}_j,
$$
where $\hat{R}_i = \big( 1,Z_i,M_i^\T,X_i^\T,\hat{L}_i^\T \big)^\T$. Recall that we have known $\lambda_n \hat{w}_j / \sqrt{n} = \lambda_n n^{-(\delta-1)/2} (\sqrt{n}\lvert \hat{\xi}_j \rvert)^{-\delta}$ tends to $\infty$. Now because
\begin{align*}
	\dfrac{ \sum_{i=1}^n \hat{R}_{i,j} \big(  Y_i - \hat{\xi}^\T \hat{R}_i \big) }{\sqrt{n}}
	&=  \dfrac{ \sum_{i=1}^n \hat{R}_{i,j} \big( \xi^\T R_i + \psi_i - \hat{\xi}^\T \hat{R}_i \big) }{\sqrt{n}} \\
	&= \dfrac{ \sum_{i=1}^n \hat{R}_{i,j}  R_i^\T \sqrt{n} (\xi - \hat{\xi}) }{n}  + \dfrac{ \sum_{i=1}^n \hat{R}_{i,j}  \sqrt{n} (R_i - \hat{R}_i)^\T \hat{\xi} }{n}  + \dfrac{\sum_{i=1}^n \hat{R}_{i,j} \psi_i}{\sqrt{n}}
\end{align*}
where the three terms all converge to some normal distribution. So we immediately know $\forall j \not\in \Tilde{\mathcal{A}}$
$$
P(j\in \hat{\mathcal{A}}_n) \leq P( 2 \sum_{i=1}^n \hat{R}_{i,j} \big(  Y_i - \hat{\xi}^\T \hat{R}_i \big) = \lambda_n \hat{w}_j ) =P( 2 \sum_{i=1}^n \hat{R}_{i,j} \big(  Y_i - \hat{\xi}^\T \hat{R}_i \big) /\sqrt{n} = \lambda_n \hat{w}_j / \sqrt{n} ) \to 0,
$$
where $2 \sum_{i=1}^n \hat{R}_{i,j} \big(  Y_i - \hat{\xi}^\T \hat{R}_i \big) /\sqrt{n} = O_p(1)$ and $\lambda_n \hat{w}_j / \sqrt{n} \to \infty$. Thus, we have the following selection consistency result
$$
\lim_{n\to \infty} P(\hat{\mathcal{A}}_n =\mathcal{A}) = 1.
$$

{\section{Proof of Corollary~\ref{COR}} \label{proof_cor}
 \textbf{Proof sketch:} 
the proof follows  from the selection consistency and standard $z$-score test. Details are provided below.

For $j\in \hat{\mathcal{A}}_n$, we have
\begin{equation}\label{eqn:p-value-decomp}
\begin{aligned}
   P(p_{n,j} \leq \alpha_j) 
   &=  P(p_{n,j} \leq \alpha_j, \hat{\mathcal{A}}_n=\mathcal{A}) + P(p_{n,j} \leq \alpha_j, \hat{\mathcal{A}}_n \neq \mathcal{A})\\
   &= P(p_{{n,j}}\leq \alpha_j,\hat{\mathcal{A}}_n=\mathcal{A}) +o(1),
\end{aligned}
\end{equation}
where the second equality holds because $P(\hat{\mathcal{A}}_n=\mathcal{A})\rightarrow 1$.
On the event $\{\hat{\mathcal{A}}_n=\mathcal{A}\}$, when $j\in \hat{\mathcal{A}}_n$, it holds that 
$j\in\mathcal A$, i.e., $\beta_{2j}\neq 0$. Consequently, $H_{0j}^c$ is equivalent to $H_{0j}$.

Note that under $H_{0j}$, $T_{n,j} \xrightarrow{d} T_j \sim N(0,1)$ as $n\to \infty$, and so $p_{n,j} = 2\{ 1- \Phi(\left\lvert T_{n,j} \right\rvert)\} \rightarrow 2\{ 1- \Phi(\left\lvert T_j \right\rvert) \} \sim U(0,1)$. This  implies that $\lim_{n\to \infty} P(p_{n,j}\leq \alpha_j) = \alpha_j$ if $H^c_{0j}$ is true. 

As discussed earlier, on the event $\{\hat{\mathcal{A}}_n=\mathcal{A}\}$, $H_{0j}^c$ is equivalent to $H_{0j}$ for $j\in \hat{\mathcal{A}}_n$. Thus, when $H^c_{0,j}$ is false, we have $\lambda_j\neq 0$. This implies $p_{n,j} = 2 \{ 1- \Phi(\lvert T_{n,j} \rvert) \} \overset{P}{\longrightarrow} 0$ for $j\in\mathcal{A}$, because $T_{n,j} = \sqrt{n}(\hat{\lambda}_j - \lambda_j) + \sqrt{n}\lambda_j / \widehat{se}(\hat{\lambda}_j) \overset{{P}}{\longrightarrow}\infty $. Then according to \eqref{eqn:p-value-decomp}, we have
$
\lim_{n\to \infty}P(p_{{n,j}}\leq \alpha_j)
=  \lim_{n\to\infty} P(p_{{n,j}}\leq \alpha_j,\hat{\mathcal{A}}_n=\mathcal{A} ) 
=1$ if $H^c_{0j}$ is false for $j\in \hat{\mathcal{A}}_n$.

To prove that $\lim_{n\to \infty} P(\hat{\mathcal{A}}_{\mathrm{act},n} = \mathcal{A}_{\mathrm{act}})\geq 1- \sum_{j\in \mathcal{A}\backslash\mathcal{A}_{\mathrm{act}}} \alpha_j$, we first show that 
\begin{align*}
\lim_{n\to \infty} P(\mathcal{A}_{\mathrm{act}}\subseteq\hat{\mathcal{A}}_{\mathrm{act},n} )=1.
\end{align*}
 By definition, for any $j\in \mathcal{A}_{\mathrm{act}}$, $H_{0j}^c$ is false. Then $P(j\in\hat{\mathcal{A}}_{\mathrm{act},n})=P(p_{n,j}\leq \alpha_j)\rightarrow 1$ according to the second result of this corollary. Then $
P(\mathcal{A}_{\mathrm{act}}\subseteq\hat{\mathcal{A}}_{\mathrm{act},n} )=P(\cap_{j\in\mathcal{A}_{\mathrm{act}}}\{j\in\hat{\mathcal{A}}_{\mathrm{act},n}\} )\rightarrow 1$. We next show that
\begin{align*}
\lim_{n\to \infty} P(\hat{\mathcal{A}}_{\mathrm{act},n}\subseteq \mathcal{A}_{\mathrm{act}})\geq 1-\sum_{j\in \mathcal{A}\backslash\mathcal{A}_{\mathrm{act}}} \alpha_j.
\end{align*}
Note that
\begin{align*}
   \lim_{n\to \infty} P(\hat{\mathcal{A}}_{\mathrm{act},n}\subseteq \mathcal{A}_{\mathrm{act}})= \lim_{n\to\infty} P(\mathcal{A}\backslash \mathcal{A}_{\mathrm{act}}\subseteq {\mathcal{A}}\backslash \hat{\mathcal{A}}_{\mathrm{act},n})= \lim_{n\to\infty} P(\mathcal{A}\backslash \mathcal{A}_{\mathrm{act}}\subseteq\hat{\mathcal{A}}_{n}\backslash \hat{\mathcal{A}}_{\mathrm{act},n}),
\end{align*}
where the second equality holds due to the selection consistency. By definition, for any $j\in \mathcal{A}\backslash \mathcal{A}_{\mathrm{act}}$, $H_{0j}^c$ is true. Then according to the first result of this corollary, we have
$P(j\in \hat{\mathcal{A}}_{n}\backslash \hat{\mathcal{A}}_{\mathrm{act},n})=1-P(p_{n,j}\leq \alpha_j)\rightarrow \alpha_j$. Consequently,
\begin{align*}
    \lim_{n\rightarrow\infty}P(\hat{\mathcal{A}}_{\mathrm{act},n}\subseteq \mathcal{A}_{\mathrm{act}})=&\lim_{n\rightarrow\infty}P(\mathcal{A}\backslash \mathcal{A}_{\mathrm{act}}\subseteq\hat{\mathcal{A}}_{n}\backslash \hat{\mathcal{A}}_{\mathrm{act},n})\\
    =&\lim_{n\rightarrow\infty}P(\cap_{j\in\mathcal{A}\backslash \mathcal{A}_{\mathrm{act}}}\{j\in\hat{\mathcal{A}}_{n}\backslash \hat{\mathcal{A}}_{\mathrm{act},n}\})\\
    =&1-\lim_{n\rightarrow\infty}P(\cup_{j\in\mathcal{A}\backslash \mathcal{A}_{\mathrm{act}}}\{j\notin\hat{\mathcal{A}}_{n}\backslash \hat{\mathcal{A}}_{\mathrm{act},n}\})\\
    \geq & 1-\sum_{j\in\mathcal{A}\backslash \mathcal{A}_{\mathrm{act}}}\lim_{n\rightarrow\infty}P(j\notin\hat{\mathcal{A}}_{n}\backslash \hat{\mathcal{A}}_{\mathrm{act},n})\\
    =& 1-\sum_{j\in\mathcal{A}\backslash \mathcal{A}_{\mathrm{act}}}\alpha_j.
\end{align*}
Combining these two pieces proves the third result of this corollary.
}

\section{Example} \label{examp}
We further illustrate  our identification strategy through the following example and highlight its difference from the null-treatment or sparsity assumptions in two related papers by \citet{miao2023identifying} and~\citet{tang2023synthetic} that focus on multi-treatment problems, which also exploit the shared confounding structure.

\begin{example} \label{exam:multitreat}
	Consider the model presented in (2.1)-(2.2), where $M\in \mathbb{R}^p$, and all other variables $Z,X,U,\eta,\varepsilon_j$ are of one-dimension, zero mean and unit variance for $j=1,\ldots,p$. Let 
	$$
	\beta_2 = ( 1, 1, 1, 1, 1, \underbrace{0, \ldots, 0}_{p-5} )^\T, ~~  \Gamma = ( 1, \ldots , 1, \underbrace{0, \ldots, 0}_{p-10} )^\T.
	$$ 
	The null treatment strategy in \citet{miao2023identifying}  assumes that the cardinality of $\mathcal{E} \cap \mathcal{F}$ should not exceed $( \lvert \mathcal{E} \rvert - t)/2$, where $\mathcal{E}$ and $\mathcal{F}$ represent index sets of confounded treatments and active treatments that have non-zero effects on the outcome respectively  in a multi-treatment scenario. Here, $t$ is the dimension of unmeasured confounding. By adapting this strategy to the context of the current mediation analysis, we find that $\lvert \mathcal{E} \cap \mathcal{F} \rvert = 5$, which is greater than $( \lvert \mathcal{E} \rvert - t )/2 = (10 -1 )/2 =4.5$. This fails to meet the null treatment assumption.
	Since  $t=1$, we deduce from Example 1 in the main manuscript that
	$
	\Delta = (1 + \Gamma^\T \Sigma_{\varepsilon}^{-1} \Gamma)^{-1}\Sigma_{\varepsilon}^{-1} \Gamma$.
	The identification assumption of the synthetic instrument approach  by \citet{tang2023synthetic} necessitates the invertibility of any $t\times t$ submatrix of $\Delta$. Here the invertibility of $\Delta$ requires that all components in $\Gamma$ are nonzero, which does not hold true in our context. In contrast, our identification assumption can be satisfied 
	 when $g_j(Z,X)$ is a nonlinear function of $Z$ and $X$ for some $j=1,\cdots,10$, as shown in the simulation studies.
\end{example}

\section{Additional simulation results} \label{add_simu}

\subsection{Simulation results for scenario 2 in Section \ref{sec:simulation}}\label{sec:simulation-scenario2}

In this section, we present the simulation results for scenario 2 described in Section~\ref{sec:simulation}. The corresponding estimation results for $\beta_2$ and the NIE are reported in Tables~\ref{tab:5+15_beta2} and~\ref{tab:5+15_NIE}, respectively. Overall, the results display patterns similar to those in Tables~\ref{tab:beta2} and~\ref{tab:NIE} for scenario~1.

\begin{table}[t]
 	\centering
 	\caption{Simulation results for estimating $\beta_2$ in scenario 2 in Section \ref{sec:simulation}. The caption details remain the same as those in Table~\ref{tab:beta2}.}
 	\label{tab:5+15_beta2}
 	\resizebox{0.95\columnwidth}{!}{
 	\begin{tabular}{ccccccccccccc}
 		\toprule
 		& & \multicolumn{3}{c}{Proposed approach} & & \multicolumn{3}{c}{Naive lasso} & & \multicolumn{3}{c}{Naive adaptive lasso}\\
 		\cmidrule(r){3-5} \cmidrule(r){7-9} \cmidrule(r){11-13}  
 		$p$ &   & 100  & 200 & 300 & & 100  & 200 & 300 & & 100  & 200 & 300\\
 		\midrule
 		& MSE    & 0.42 & 0.17  & 0.25  &  & 0.36  &  0.44 & 0.52 & & 0.30  & 0.30  & 0.53 \\
		$n=300,\varphi = 1 $	& TP    & 19.91 & 20  & 20  &  & 20  &  20 & 20 & & 20  & 20  & 20 \\
		& FP    & 0.27 & 0.04  & 0.02  &  & 33.07  &  47.50 & 54.18 & & 0  & 0  & 0\\
		\midrule
		& MSE   & 0.60 & 0.41  & 0.45 &  & 1.99  &  2.15 & 2.31 & & 1.68  & 1.73  & 4.63  \\
		$n=300,\varphi = 4 $	& TP   & 19.95 & 20  & 19.98  &  & 20  &  20 & 20 & & 20  & 20  & 19.80 \\
		& FP   & 1.66 & 3.26  & 0.18  &  & 32.61  &  46.09 & 55.44 & & 5.49  & 5.97  & 1.46 \\
		\midrule
		& MSE    & 0.12 & 0.11  & 0.12  &  & 0.23  &  0.25 & 0.27 & & 0.24  & 0.24  & 0.25 \\
		$n=600,\varphi = 1 $	& TP    & 20 & 20  & 20  &  & 20  &  20 & 20 & & 20  & 20  & 20 \\
		& FP   & 0.02 & 0  & 0.01  &  & 23.11  &  47.64 & 54.91 & & 0  & 0  & 0 \\
		\midrule
		& MSE  & 0.27 & 0.19  & 0.19  &  & 1.65  &  1.73 & 1.78 & & 1.50  & 1.51  & 1.50\\
		$n=600,\varphi = 4 $	& TP   & 19.98 & 20  & 20  &  & 20  &  20 & 20 & & 20  & 20  & 20 \\
		& FP    & 0.17 & 0.02  & 0.02  &  & 32.53  &  45.25 & 53.11 & & 4.95  & 4.98  & 4.99 \\
		\midrule
		& MSE    & 0.09 & 0.09  & 0.09  &  & 0.21  &  0.21 & 0.21 & & 0.22  & 0.22  & 0.22 \\
		$n=1000,\varphi = 1 $	& TP    & 20 & 20  & 20  &  & 20  &  20 & 20 & & 20  & 20  & 20 \\
		& FP   & 0 & 0  & 0  &  & 9.91  &  26.50 & 45.08 & & 0  & 0  & 0     \\
		\midrule
		& MSE     & 0.14 & 0.16  & 0.14  &  & 1.55  &  1.59 & 1.62 & & 1.44  & 1.44  & 1.44\\
		$n=1000,\varphi = 4 $ & TP   & 20 & 20  & 20  &  & 20  &  20 & 20 & & 20  & 20  & 20 \\
		& FP & 0.03 & 0.04  & 0  &  & 31.24  &  45.98 & 56.98 & & 5  & 5  & 5  \\
 		\bottomrule
 	\end{tabular}%
 	}
 \end{table}

\begin{table}[t]
 	\centering
 	\caption{
    {Simulation results for estimating $\mathrm{NIE}$ in scenario 2 in Section \ref{sec:simulation}. The caption details remain the same as those in Table~\ref{tab:NIE}.}}
 	\label{tab:5+15_NIE}
 	\resizebox{0.95\columnwidth}{!}{
 	\begin{tabular}{ccccccccccccc}
 		\toprule
 		& & \multicolumn{3}{c}{Proposed approach} & & \multicolumn{3}{c}{Naive lasso} & & \multicolumn{3}{c}{Naive adaptive lasso}\\
 		\cmidrule(r){3-5} \cmidrule(r){7-9} \cmidrule(r){11-13}  
 		$p$ &   & 100  & 200 & 300 & & 100  & 200 & 300 & & 100  & 200 & 300\\
 		\midrule
 		& MSE    & 0.73 & 0.47  & 0.52  &  & 0.71  &  0.74 & 0.82 & & 0.64  & 0.64  & 0.85 \\
		$n=300,\varphi = 1 $	& TP    & 19.91 & 20  & 20  &  & 20  &  20 & 20 & & 20  & 20  & 20 \\
		& FP    & 0.27 & 0.04  & 0.02  &  & 33.07  &  47.50 & 54.18 & & 0  & 0  & 0\\
		\midrule
		& MSE   & 0.93 & 0.73  & 0.73 &  & 2.45  &  2.52 & 2.74 & & 2.09  & 2.11  & 5.05  \\
		$n=300,\varphi = 4 $	& TP   & 19.95 & 20  & 19.98  &  & 20  &  20 & 20 & & 20  & 20  & 19.80 \\
		& FP   & 1.66 & 3.26  & 0.18  &  & 32.61  &  46.09 & 55.44 & & 5.49  & 5.97  & 1.46 \\
		\midrule
		& MSE    & 0.27 & 0.25  & 0.26  &  & 0.40  &  0.41 & 0.42 & & 0.41  & 0.43  & 0.43 \\
		$n=600,\varphi = 1 $	& TP    & 20 & 20  & 20  &  & 20  &  20 & 20 & & 20  & 20  & 20 \\
		& FP   & 0.02 & 0  & 0.01  &  & 23.11  &  47.64 & 54.91 & & 0  & 0  & 0 \\
		\midrule
		& MSE  & 0.43 & 0.33  & 0.34  &  & 1.85  &  1.95 & 1.96 & & 1.69  & 1.79  & 1.74\\
		$n=600,\varphi = 4 $	& TP   & 19.98 & 20  & 20  &  & 20  &  20 & 20 & & 20  & 20  & 20 \\
		& FP    & 0.17 & 0.02  & 0.02  &  & 32.53  &  45.25 & 53.11 & & 4.95  & 4.98  & 4.99 \\
		\midrule
		& MSE    & 0.19 & 0.18  & 0.18  &  & 0.31  &  0.32 & 0.31 & & 0.34  & 0.35  & 0.32 \\
		$n=1000,\varphi = 1 $	& TP    & 20 & 20  & 20  &  & 20  &  20 & 20 & & 20  & 20  & 20 \\
		& FP   & 0 & 0  & 0  &  & 9.91  &  26.50 & 45.08 & & 0  & 0  & 0     \\
		\midrule
		& MSE     & 0.24 & 0.26  & 0.23  &  & 1.69  &  1.76 & 1.72 & & 1.63  & 1.64  & 1.56\\
		$n=1000,\varphi = 4 $ & TP   & 20 & 20  & 20  &  & 20  &  20 & 20 & & 20  & 20  & 20 \\
		& FP & 0.03 & 0.04  & 0  &  & 31.24  &  45.98 & 56.98 & & 5  & 5  & 5  \\
 		\bottomrule
 	\end{tabular}%
 	}
 \end{table}
{
\subsection{Simulation in the sparse exposure setting}\label{sec:sim-sparse-exposure}

To evaluate the performance of our estimator when the exposure affects only a small subset of mediators, we consider a setting in which only the first ten mediators are influenced by the exposure. Relative to the simulation design in Section \ref{sec:simulation} of the main text, the only modification is that we specify the exposure–mediator effects as
$\gamma_1 = (\underbrace{1,\ldots,1}_{10},0,\ldots,0)^\T$,
while the data-generating models and all other parameters remain identical to those in the main context. Under this setting, the estimation results for $\beta_2$ under the two distinct scenarios are reported in Tables~\ref{tab:beta2_sparse_exposure} and~\ref{tab:5+15_beta2_sparse_exposure}. The corresponding results for the NIE are presented in Tables~\ref{tab:NIE_sparse_exposure} and~\ref{tab:5+15_NIE_sparse_exposure}. Overall, the performance is highly consistent with that observed in Section~\ref{sec:simulation} of the main text (see Tables~\ref{tab:beta2}, \ref{tab:NIE}, \ref{tab:5+15_beta2}, and \ref{tab:5+15_NIE}).
}

\begin{table}[t]
	\centering
	\caption{ {Simulation results  for estimating $\beta_2$ in scenario 1 under the sparse exposure setting. The caption details remain the same as those in Table \ref{tab:beta2}.}}
	\label{tab:beta2_sparse_exposure}
	\resizebox{0.95\columnwidth}{!}{
	\begin{tabular}{ccccccccccccc}
		\toprule
		& & \multicolumn{3}{c}{Proposed approach} & & \multicolumn{3}{c}{Naive lasso} & & \multicolumn{3}{c}{Naive adaptive lasso}\\
		\cmidrule(r){3-5} \cmidrule(r){7-9} \cmidrule(r){11-13}  
		$p$ &   & 100  & 200 & 300 & & 100  & 200 & 300 & & 100  & 200 & 300\\
		\midrule
		& MSE    & 0.04 & 0.03  & 0.04  &  & 0.12  &  0.11 & 0.12 & & 0.15  & 0.15  & 0.15 \\
		$n=300,\varphi = 1 $	& TP    & 5 & 5  & 5  &  & 5  &  5 & 5 & & 5  & 5  & 5 \\
		& FP    & 0.32 & 0.12  & 0.01  &  & 11.98  &  14.45 & 16.87 & & 0  & 0  & 0\\
		\midrule
		& MSE   & 0.22 & 0.34  & 0.10 &  & 1.39  & 1.37 & 1.38 & & 1.49  & 1.51  & 2.06  \\
		$n=300,\varphi = 4 $	& TP   & 5 & 5  & 4.99  &  & 5  &  5 & 5 & & 5  & 5  & 5 \\
		& FP   & 3.48 & 8.86  & 0.38  &  & 11.54  & 14.03 & 16.59 & & 5.33  & 5.68  & 1.36 \\
		\midrule
		& MSE    & 0.02 & 0.02  & 0.02  &  & 0.10  &  0.10 & 0.10 & & 0.14  & 0.14  & 0.14 \\
		$n=600,\varphi = 1 $	& TP    & 5 & 5  & 5  &  & 5  &  5 & 5 & & 5  & 5  & 5 \\
		& FP   & 0 & 0  & 0  &  & 13.12  &  14.74 & 15.93 & & 0  & 0 & 0 \\
		\midrule
		& MSE  & 0.06 & 0.04  & 0.04  &  & 1.36  & 1.36 & 1.36 & & 1.41  & 1.42  & 1.42 \\
		$n=600,\varphi = 4 $	& TP   & 5 & 5  & 5  &  & 5  &  5 & 5 & & 5  & 5  & 5 \\
		& FP    & 0.17 & 0  & 0  &  & 11.98  &  14.48 & 15.66 & & 5  & 4.99  & 5 \\
		\midrule
		& MSE    & 0.01 & 0.01  & 0.01  &  & 0.09  &  0.09 & 0.09 & & 0.13  & 0.13  & 0.13 \\
		$n=1000,\varphi = 1 $	& TP    & 5 & 5  & 5  &  & 5  &  5 & 5 & & 5  & 5  & 5 \\
		& FP   & 0 & 0  & 0  &  & 12.07  &  15.26 & 16.54 & & 0  & 0  & 0     \\
		\midrule
		& MSE     & 0.03 & 0.03  & 0.03  &  & 1.37  &  1.36 & 1.36 & & 1.40  & 1.39  & 1.39\\
		$n=1000,\varphi = 4 $ & TP   & 5 & 5  & 5  &  & 5  &  5 & 5 & & 5  & 5  & 5 \\
		& FP & 0.01 & 0  & 0  &  & 12.18  &  14.24 & 15.85 & & 5  & 5  & 5  \\
		\bottomrule
	\end{tabular}%
	}
\end{table}%

 \begin{table}[t]
 	\centering
 	\caption{
    {Simulation results for estimating $\beta_2$ in scenario 2 under the sparse exposure setting. The  caption details remain the same as those in Table~\ref{tab:5+15_beta2}.}}
 	\label{tab:5+15_beta2_sparse_exposure}
 	\resizebox{0.95\columnwidth}{!}{
 	\begin{tabular}{ccccccccccccc}
 		\toprule
 		& & \multicolumn{3}{c}{Proposed approach} & & \multicolumn{3}{c}{Naive lasso} & & \multicolumn{3}{c}{Naive adaptive lasso}\\
 		\cmidrule(r){3-5} \cmidrule(r){7-9} \cmidrule(r){11-13}  
 		$p$ &   & 100  & 200 & 300 & & 100  & 200 & 300 & & 100  & 200 & 300\\
 		\midrule
 		& MSE    & 0.35 & 0.15  & 0.21  &  & 0.35  &  0.42 & 0.49 & & 0.28  & 0.28  & 0.48 \\
		$n=300,\varphi = 1 $	& TP    & 19.94 & 20  & 20  &  & 20  &  20 & 20 & & 20  & 20  & 20 \\
		& FP    & 0.31 & 0.03  & 0.01  &  & 32.13  &  47.19 & 54.93 & & 0  & 0  & 0 \\
		\midrule
		& MSE   & 0.56 & 0.43  & 0.42 &  & 1.97  &  2.12 & 2.28 & & 1.68  & 1.73  & 4.33  \\
		$n=300,\varphi = 4 $	& TP   & 19.97 & 20  & 19.98  &  & 20  &  20 & 20 & & 20  & 20  & 19.85 \\
		& FP   & 2.03 & 4.77  & 0.21  &  & 32.2  &  45.87 & 55.65 & & 5.65  & 6.46  & 1.45 \\
		\midrule
		& MSE    & 0.10 & 0.10  & 0.10  &  & 0.21  &  0.24 & 0.26 & & 0.22  & 0.23  & 0.23 \\
		$n=600,\varphi = 1 $	& TP    & 20 & 20  & 20  &  & 20  & 20 & 20 & & 20  & 20  & 20 \\
		& FP   & 0.01 & 0  & 0  &  & 29.6  &  47.52 & 53.19 & & 0  & 0 & 0 \\
		\midrule
		& MSE  & 0.24 & 0.17  & 0.17  &  & 1.64  & 1.71 & 1.76 & & 1.50  & 1.50 & 1.50 \\
		$n=600,\varphi = 4 $	& TP   & 20 & 20  & 20  &  & 20  &  20 & 20 & & 20  & 20  & 20 \\
		& FP    & 0.15 & 0.02  & 0.01  &  & 31.95  & 44.84 & 51.5 & & 4.96  & 4.98  & 4.99 \\
		\midrule
		& MSE    & 0.07 & 0.07  & 0.07  &  & 0.18  &  0.18 & 0.19 & & 0.20  & 0.20  & 0.20 \\
		$n=1000,\varphi = 1 $	& TP    & 20 & 20  & 20  &  & 20  &  20 & 20 & & 20  & 20  & 20 \\
		& FP   & 0 & 0  & 0  &  & 14.24  &  42.95 & 55.73 & & 0  & 0  & 0     \\
		\midrule
		& MSE     & 0.12 & 0.12  & 0.12  &  & 1.54  &  1.58 & 1.60 & & 1.44  & 1.44  & 1.44\\
		$n=1000,\varphi = 4 $ & TP   & 20 & 20  & 20  &  & 20  &  20 & 20 & & 20  & 20  & 20 \\
		& FP & 0.02 & 0.03  & 0  &  & 31.66  &  45.01 & 55.01 & & 5  & 5  & 5  \\
 		\bottomrule
 	\end{tabular}%
 	}
 \end{table}

\begin{table}[t]
	\centering
	\caption{ {Simulation results  for estimating $\mathrm{NIE}$ in scenario 1 under the sparse exposure setting.  The  caption details remain the same as those in Table~\ref{tab:NIE}.}}
	\label{tab:NIE_sparse_exposure}
	\resizebox{0.95\columnwidth}{!}{
	\begin{tabular}{ccccccccccccc}
		\toprule
		& & \multicolumn{3}{c}{Proposed approach} & & \multicolumn{3}{c}{Naive lasso} & & \multicolumn{3}{c}{Naive adaptive lasso}\\
		\cmidrule(r){3-5} \cmidrule(r){7-9} \cmidrule(r){11-13}  
		$p$ &   & 100  & 200 & 300 & & 100  & 200 & 300 & & 100  & 200 & 300\\
		\midrule
		& MSE    & 0.17 & 0.15 & 0.14  &  & 0.27  &  0.23 & 0.25 & & 0.30  & 0.31  & 0.32 \\
		$n=300,\varphi = 1 $	& TP    & 5 & 5 & 5  &  & 5  & 5 & 5 & & 5  & 5  & 5 \\
		& FP    & 0.32 & 0.12 & 0.01  &  & 11.98  & 14.45 & 16.87 & & 0  & 0  & 0 \\
		\midrule
		& MSE   & 0.29 & 0.24 & 0.21  &  & 1.64  &  1.55 & 1.62 & & 1.68  & 1.68  & 2.36  \\
		$n=300,\varphi = 4 $	& TP   & 5 & 5 & 5  &  & 5  &  5 & 5 & & 5  & 5  & 5 \\
		& FP   & 3.48 & 8.86 & 0.38  &  & 11.54  & 14.03 & 16.59 & & 5.33  & 5.68  & 1.36 \\
		\midrule
		& MSE    & 0.08 & 0.07 & 0.08  &  & 0.16  &  0.16 & 0.16 & & 0.22  & 0.24  & 0.23 \\
		$n=600,\varphi = 1 $	& TP    & 5 & 5 & 5  &  & 5  &  5 & 5 & & 5  & 5  & 5 \\
		& FP   & 0 & 0 & 0  &  & 13.12  & 14.74 & 15.93 & & 0  & 0  & 0 \\
		\midrule
		& MSE  & 0.11 & 0.10 & 0.11  &  & 1.45  &  1.48 & 1.45  & & 1.51  & 1.59  & 1.56 \\
		$n=600,\varphi = 4 $	& TP   & 5 & 5 & 5  &  & 5  & 5 & 5 & & 5  & 5  & 5 \\
		& FP    & 0.17 & 0 & 0  &  & 11.98  & 14.48 & 15.66 & & 5  & 5  & 5 \\
		\midrule
		& MSE    & 0.06 & 0.05 & 0.05  &  & 0.14  &  0.14 & 0.13 & & 0.20  & 0.20  & 0.18 \\
		$n=1000,\varphi = 1 $	& TP    & 5 & 5 & 5  &  & 5  &  5 &5 & & 5  & 5  &5 \\
		& FP   & 0 & 0 & 0  &  & 12.07  &  15.26 & 16.54 & & 0  & 0  & 0     \\
		\midrule
		& MSE     & 0.07 & 0.07 & 0.07  &  & 1.45  &  1.47 & 1.40 & & 1.53  & 1.53  & 1.46 \\
		$n=1000,\varphi = 4 $ & TP   & 5 & 5 & 5  &  & 5 & 5 &5 & & 5  &5  &5 \\
		& FP & 0.01 & 0 & 0  &  & 12.18  &  14.24 &15.85 & & 5  & 5  & 5  \\
		\bottomrule
	\end{tabular}%
	}
\end{table}%

 \begin{table}[t]
 	\centering
 	\caption{
    {Simulation results for estimating $\mathrm{NIE}$ in scenario 2 under the sparse exposure setting. The  caption details remain the same as those in Table~\ref{tab:5+15_NIE}.}}
 	\label{tab:5+15_NIE_sparse_exposure}
 	\resizebox{0.95\columnwidth}{!}{
 	\begin{tabular}{ccccccccccccc}
 		\toprule
 		& & \multicolumn{3}{c}{Proposed approach} & & \multicolumn{3}{c}{Naive lasso} & & \multicolumn{3}{c}{Naive adaptive lasso}\\
 		\cmidrule(r){3-5} \cmidrule(r){7-9} \cmidrule(r){11-13}  
 		$p$ &   & 100  & 200 & 300 & & 100  & 200 & 300 & & 100  & 200 & 300\\
 		\midrule
 		& MSE    & 0.56 & 0.36 & 0.35  &  & 0.46  &  0.42 & 0.43 & & 0.49  & 0.49  & 0.49 \\
		$n=300,\varphi = 1 $	& TP    & 19.94 & 20 & 20  &  & 20  & 20 & 20 & & 20  &20  & 20 \\
		& FP    & 0.31 & 0.03 & 0.01  &  & 32.13  & 47.19 & 54.93 & & 0  & 0  & 0 \\
		\midrule
		& MSE   & 0.64 & 0.43 & 0.45  &  & 1.89  & 1.77 & 1.84 & & 1.88  &1.89  & 2.50  \\
		$n=300,\varphi = 4 $	& TP   & 19.97 & 20 & 19.98  &  & 20  & 20 & 20 & & 20  &20  & 20 \\
		& FP   & 2.03 & 4.77 & 0.21  &  & 32.20  & 45.87 & 55.65 & & 5.65  & 6.46 & 16.3 \\
		\midrule
		& MSE    & 0.19 & 0.18 & 0.18  &  & 0.26  & 0.26 & 0.25 & & 0.32  & 0.34  & 0.32 \\
		$n=600,\varphi = 1 $	& TP    & 20 & 20 & 20  &  & 20  & 20 & 20 & & 20  & 20  & 20 \\
		& FP   & 0.01 & 0 & 0  &  & 29.6  & 47.52 & 53.19 & & 0  & 0  & 0 \\
		\midrule
		& MSE  & 0.29 & 0.21 & 0.21  &  & 1.58  & 1.61 &1.57 & & 1.61  & 1.70  & 1.66 \\
		$n=600,\varphi = 4 $	& TP   & 20 & 20 & 20  &  & 20  & 20 & 20 & & 20  & 20  & 20 \\
		& FP    & 0.15 & 0.02 & 0.01  &  & 31.95  & 44.84 & 51.50 & & 4.96  &4.98  & 4.99 \\
		\midrule
		& MSE    & 0.13 & 0.12 & 0.12  &  & 0.19  &  0.20 & 0.19 & & 0.26  & 0.26  & 0.24 \\
		$n=1000,\varphi = 1 $	& TP    &20 & 20 & 20  &  &20 & 20 & 20 & &20 & 20  & 20 \\
		& FP   & 0 & 0 & 0  &  &14.24  & 42.95 &55.73 & & 0  & 0  & 0     \\
		\midrule
		& MSE     & 0.14 & 0.14 & 0.13  &  &1.54  &1.55 &1.48 & &1.57  &1.58  &1.51 \\
		$n=1000,\varphi = 4 $ & TP   & 20 & 20 & 20  &  & 20  &20 & 20 & & 20  &20  & 20 \\
		& FP & 0.02 & 0.03 & 0  &  &31.66  &45.01 &55.01 & & 5  & 5  & 5  \\
 		\bottomrule
 	\end{tabular}%
 	}
 \end{table}

 {
 \subsection{Simulation with misspecified outcome model}\label{sec:sim-misspecify}
 To assess the robustness of our proposed estimator against model misspecification, we extend the analysis to generate the outcome $Y$ in the following new way:
\begin{align}
Y= \beta_1 Z + \beta_2^\T M + \beta_3 X + \varphi_1 \sin(X) + \varphi U + \varphi_1 U^2 + \eta.
\label{modelforRobust_sinX}
\end{align}
 Here, $\beta_1$, $\beta_2$, and $\beta_3$ take the same values as those specified in the correctly specified model setting in Section \ref{sec:simulation}. We consider a relatively large sample size $n=1000$ and relatively strong unmeasured confounding with $\varphi=4$. The parameter $\varphi_1$ varies across $\{ 0.5, 1, 1.5, 2 \}$, and the results are presented in Tables~\ref{tab:beta2_1_robust_sinX} and~\ref{tab:beta2_2_robust_sinX} for the two distinct scenarios of $\beta_2$. 
Table~\ref{tab:beta2_1_robust_sinX} for scenario 1 shows that our approach consistently outperforms the two naive approaches in terms of MSE, regardless of the specific value of $\varphi_1$. When the outcome model exhibits mild misspecification, with lower $\varphi_1$ values indicating less deviation, our approach can precisely exclude false signals and exhibit minimal false positives. However, in the presence of substantial misspecification, characterized by $\varphi_1=2$, all three methods are unable to completely avoid inducing false signals. Similar results  for scenario 2 are presented in Table~\ref{tab:beta2_2_robust_sinX}. The corresponding results for estimating NIE for the two different scenarios in this new setting are provided in Tables \ref{tab: NIE_1_robust_sinX} and \ref{tab:NIE_2_robust_sinX}, respectively.
Overall, these findings imply that our proposed estimator can still provide good performance  when the outcome model is moderately misspecified.
}

\begin{table}
	\centering
	\caption{{Simulation results for estimating $\beta_2$ in scenario 1 under  model~\eqref{modelforRobust_sinX}. The parameter $\varphi_1$ indicates the extent of model misspecification, and
		the rest of the caption details
		remain the same as those in Table~\ref{tab:beta2}.
	}}
	\label{tab:beta2_1_robust_sinX}
	\resizebox{0.95\columnwidth}{!}{
	\begin{tabular}{ccccccccccccc}
		\toprule
		& & \multicolumn{3}{c}{Proposed approach} & & \multicolumn{3}{c}{Naive lasso} & & \multicolumn{3}{c}{Naive adaptive lasso}\\ 
		\cmidrule(r){3-5} \cmidrule(r){7-9} \cmidrule(r){11-13}  
		$p$ &   & 100  & 200 & 300 & & 100  & 200 & 300 & & 100  & 200 & 300\\ 	
		\midrule
		& MSE    & 0.04  & 0.04  & 0.03  &  & 1.39  &  1.38 & 1.38 & & 1.40  & 1.39  & 1.39 \\ 
		$\varphi_1 = 0.5 $	& TP   & 5  & 5  & 5  &  & 5  &  5 & 5 & & 5  & 5  & 5 \\ 
		& FP    & 0.01  & 0.01  & 0  &  & 11.22  &  13.71 & 15.19 & & 5  & 5  & 5 \\ 
		\midrule
		& MSE     & 0.06  & 0.05  & 0.05  &  & 1.42  &  1.41 & 1.41 & & 1.40  & 1.39  & 1.39\\ 
		$\varphi_1 = 1 $ 	& TP   & 5  & 5  & 5  &  & 5  &  5 & 5 & & 5  & 5  & 5 \\ 
		& FP  & 0.01  & 0.02  & 0  &  & 11.38  &  14.52 & 14.21 & & 5  & 5  & 5  \\ 
		\midrule
		& MSE  & 0.09  & 0.12  & 0.10  &  & 1.45  &  1.45 & 1.44 & & 1.40  & 1.39  & 1.39 \\ 
		$\varphi_1 = 1.5 $	& TP   & 5  & 5  & 5  &  & 5  &  5 & 5 & & 5  & 5  & 5 \\ 
		& FP    & 0.26  & 1.16  & 0.94  &  & 11.22  &  14.45 & 14.44 & & 5  & 5  & 5 \\ 
		\midrule
		& MSE    & 0.22  & 0.42  & 0.36  &  & 1.50  &  1.50 & 1.49 & & 1.40  & 1.39  & 1.39 \\ 
		$\varphi_1 = 2 $	& TP   & 5  & 5  & 5  &  & 5  &  5 & 5 & & 5  & 5  & 5 \\ 
		& FP   & 2.65  & 7.6  & 6.12  &  & 11.15  &  14.48 & 14.48 & & 5  & 5  & 5  \\ 
		\bottomrule
	\end{tabular}%
	}
\end{table}%

\begin{table}
	\centering
	\caption{{Simulation results for estimating $\beta_2$ in scenario 2 under  model~\eqref{modelforRobust_sinX}. The parameter $\varphi_1$ indicates the extent of model misspecification, and
		the rest of the caption details
		remain the same as those in Table~\ref{tab:5+15_beta2}.
	}}
	\label{tab:beta2_2_robust_sinX}
	\resizebox{0.95\columnwidth}{!}{
	\begin{tabular}{ccccccccccccc}
		\toprule
		& & \multicolumn{3}{c}{Proposed approach} & & \multicolumn{3}{c}{Naive lasso} & & \multicolumn{3}{c}{Naive adaptive lasso}\\
		\cmidrule(r){3-5} \cmidrule(r){7-9} \cmidrule(r){11-13}  
		$p$ &   & 100  & 200 & 300 &  & 100  & 200 & 300 & & 100  & 200 & 300\\	
		\midrule
		& MSE    & 0.15  & 0.19  & 0.15  &  & 1.58  &  1.62 & 1.66 & & 1.44  & 1.44  & 1.44 \\ 
		$\varphi_1 = 0.5 $	& TP   & 20  & 20  & 20  &  & 20  &  20 & 20 & & 20  & 20  & 20 \\ 
		& FP    & 0.01  & 0.03  & 0  &  & 31.9  &  46.08 & 55.39 & & 5  & 5  & 5 \\ 
		\midrule
		& MSE     & 0.19  & 0.22  & 0.18  &  & 1.65  &  1.73 & 1.77 & & 1.44  & 1.44  & 1.44\\ 
		$\varphi_1 = 1 $ 	& TP   & 20  & 20  & 20  &  & 20  &  20 & 20 & & 20  & 20  & 20 \\ 
		& FP  & 0.02  & 0.03  & 0  &  & 31.37  &  45.43 & 54.07 & & 5  & 5  & 5  \\ 
		\midrule
		& MSE  & 0.24  & 0.28  & 0.25  &  & 1.78  &  1.90 & 1.96 & & 1.44  & 1.44  & 1.44 \\ 
		$\varphi_1 = 1.5 $	& TP   & 20  & 20  & 20  &  & 20  &  20 & 20 & & 20  & 20  & 20 \\ 
		& FP    & 0.06  & 0.06  & 0.23  &  & 30.64  &  44.37 & 53.01 & & 5  & 5  & 5 \\ 
		\midrule
		& MSE    & 0.33  & 0.41  & 0.42  &  & 1.96  &  2.13 & 2.23 & & 1.44  & 1.44  & 1.44 \\ 
		$\varphi_1 = 2 $	& TP   & 20  & 20  & 20  &  & 20  &  20 & 20 & & 20  & 20  & 20 \\ 
		& FP   & 0.21  & 1.88  & 2.74  &  & 30.29  &  44.02 & 53.09 & & 5  & 5  & 5  \\ 
		\bottomrule
	\end{tabular}%
	}
\end{table}%

\begin{table}
	\centering
	\caption{{Simulation results for estimating $\mathrm{NIE}$ in scenario 1 under  model~\eqref{modelforRobust_sinX}. The parameter $\varphi_1$ indicates the extent of model misspecification, and
		the rest of the caption details
		remain the same as those in Table~\ref{tab:NIE}.
	}}
	\label{tab: NIE_1_robust_sinX}
	\resizebox{0.95\columnwidth}{!}{
	\begin{tabular}{ccccccccccccc}
		\toprule
		& & \multicolumn{3}{c}{Proposed approach} & & \multicolumn{3}{c}{Naive lasso} & & \multicolumn{3}{c}{Naive adaptive lasso}\\
		\cmidrule(r){3-5} \cmidrule(r){7-9} \cmidrule(r){11-13}  
		$p$ &   & 100  & 200 & 300 &  & 100  & 200 & 300 & & 100  & 200 & 300\\	
		\midrule
		& MSE    & 0.08  & 0.08  & 0.07  &  & 1.47  & 1.50 & 1.44 & & 1.22  & 1.47  & 1.12 \\ 
		$\varphi_1 = 0.5 $	& TP    & 5  &  5 & 5  &  & 5  & 5  &  5 & &  5  & 5  & 5 \\ 
		& FP    & 0.01  & 0.01  & 0   & & 11.22  & 13.71 &  15.19 & & 5 &  5 & 5 \\ 
		\midrule
		& MSE     & 0.09  & 0.09  & 0.09  & & 1.49  & 1.53  & 1.46  & & 1.22  & 1.47  & 1.12 \\ 
		$\varphi_1 = 1 $ 	& TP   & 5  & 5  & 5  & & 5  & 5   & 5 & & 5  & 5  & 5 \\ 
		& FP  & 0.01 & 0.02  & 0  &  & 11.38 & 14.52  & 14.21 & & 5  & 5  & 5  \\ 
		\midrule
		& MSE  & 0.12  & 0.16  & 0.14  &  & 1.53  & 1.56 &  1.49 &  & 1.22 & 1.47  & 1.12 \\ 
		$\varphi_1 = 1.5 $	& TP   & 5  & 5  & 5  &  & 5  & 5  & 5 & & 5 & 5  & 5 \\ 
		& FP    & 0.26 & 1.16  & 0.94  & & 11.22  & 14.45  & 14.44 & & 5  & 5  & 5 \\ 
		\midrule
		& MSE    & 0.25  & 0.46  & 0.40  & & 1.57  & 1.61  & 1.54 & & 
		1.22  & 1.47 &  1.12 \\ 
		$\varphi_1 = 2 $	& TP    & 5  & 5  & 5 &  & 5  & 5   & 5 & & 5  &  5 &  5 \\ 
		& FP   & 2.65  & 7.6  & 6.12  &  & 11.15 & 14.48  & 14.48 & & 5  & 5 &  5  \\ 
		\bottomrule
	\end{tabular}%
	}
\end{table}%

\begin{table}
	\centering
	\caption{{Simulation results for estimating $\mathrm{NIE}$ in scenario 2 under  model~\eqref{modelforRobust_sinX}. The parameter $\varphi_1$ indicates the extent of model misspecification, and
		the rest of the caption details
		remain the same as those in Table~\ref{tab:5+15_NIE}.
	}}
	\label{tab:NIE_2_robust_sinX}
	\resizebox{0.95\columnwidth}{!}{
	\begin{tabular}{ccccccccccccc}
		\toprule
		& & \multicolumn{3}{c}{Proposed approach} & & \multicolumn{3}{c}{Naive lasso} & & \multicolumn{3}{c}{Naive adaptive lasso}\\ 
		\cmidrule(r){3-5} \cmidrule(r){7-9} \cmidrule(r){11-13}  
		$p$ &   & 100  & 200 & 300 & & 100  & 200 & 300 & & 100  & 200 & 300\\ 	
		\midrule
		& MSE    & 0.25  & 0.28  & 0.24  &  & 1.71  & 1.79 & 1.76 & & 1.33  & 1.56 & 1.22 \\ 
		$\varphi_1 = 0.5 $	& TP    & 20  &  20 & 20  &  & 20  & 20  & 20 & &  20  & 20  & 20 \\ 
		& FP    & 0.01  & 0.03  & 0   & & 31.9  & 46.08 & 55.39 & & 5 &  5 & 5 \\ 
		\midrule
		& MSE     & 0.28  & 0.32  & 0.28  & & 1.78  & 1.90  & 1.87  & & 1.33  & 1.56  & 1.22 \\ 
		$\varphi_1 = 1 $ 	& TP   & 20  & 20  & 20  & & 20  & 20  & 20 & & 20  & 20  & 20 \\ 
		& FP  & 0.02 & 0.03  & 0  &  & 31.37 & 45.43  & 54.07 & & 5  & 5  & 5  \\ 
		\midrule
		& MSE  & 0.33  & 0.38  & 0.34  &  & 1.91  & 2.06 & 2.05 &  & 1.33 & 1.56  & 1.22 \\ 
		$\varphi_1 = 1.5 $	& TP   & 20  & 20  & 20  &  & 20  & 20  & 20 & & 20 & 20  & 20 \\ 
		& FP    & 0.06  & 0.06  & 0.23  & & 30.64  & 44.37  & 53.01 & & 5 & 5  & 5 \\ 
		\midrule
		& MSE    & 0.42  & 0.51  & 0.52  & & 2.08 & 2.30  & 2.32 & & 
		1.33  & 1.56 & 1.22 \\ 
		$\varphi_1 = 2 $	& TP    & 20  & 20  & 20 &  & 20  & 20   & 20 & & 20  &  20 & 20 \\ 
		& FP   & 0.21  & 1.88  & 2.74  &  & 30.29 & 44.02  & 53.09 & & 5  & 5  &  5  \\ 
		\bottomrule
	\end{tabular}%
	}
\end{table}%


\section{Extension to nonlinear outcome model setting}\label{app:nonlinear}
As illustrated in the main text, our identification conclusions hold under the linear outcome model setting. Here, we show that the extension to nonlinear outcome model is also feasible. Consider the following model:
\begin{align}
	Y &=  a_1(Z,X;b_1) + \sum_{h=1}^{w_2}  b_{2h} a_{2h}(Z,X) M + \varphi^{\T} U + \eta, \label{eqn:modelforY_extension}\\
	\check{M} &= \Gamma U + \varepsilon. \label{eqn:modelforM_extension}
\end{align}
where $\check{M} = M - g(Z,X)$ with unknown function $g(\cdot)$ and $ a_1(\cdot), \{ a_{2h}(\cdot) \}_{h=1}^{w_2}$ are known linearly independent functions. Here, $(Z,X) \indep (U,\varepsilon,\eta)$ and $U,\varepsilon,\eta$ are mutually uncorrelated. Without loss of generality, we assume $E(U) = 0_t$, $\mathrm{cov}(U) = I_t$. Consistent with the main paper, we also constrain $\Sigma_{\varepsilon}=\mathrm{cov}(\varepsilon)$ in \eqref{eqn:modelforM_extension} to be diagonal following the standard latent factor model assumption \citep{anderson1956statistical}. We establish the identifiability results in the following theorem.

\begin{theorem} \label{THM:IDENTIFICATION_1_extension}
    Under model~\eqref{eqn:modelforY_extension}-\eqref{eqn:modelforM_extension}, $ b_1$ and $\{ b_{2h} \}_{h=1}^{ w_2}$ are identifiable if the following conditions hold:
	\begin{itemize}
		\item[(i)] after deleting any row of $\Gamma$, there remain two disjoint sub-matrices of full column rank;
		
		\item[(ii)] The following estimating equations have a unique solution:
  $$
  E\bigg[  \{ Y - a_1(Z,X;b_1) -b_2^\T a_2(Z,X,M) - \varphi^\T L \} \cdot \bigg\{ \dfrac{\partial a_1(Z,X;b_1)}{\partial b_1^\T} ,a_2(Z,X,M)^\T ,L^\T \bigg\}  \bigg] =0,
  $$
	\end{itemize}
 where 
 \begin{gather*}
 L = \Delta^{\T} \{ M -  g(Z,X;\gamma) \},~~ \Delta = (\Gamma \Gamma^{\T} + \Sigma_{\varepsilon} )^{-1} \Gamma, ~~ b_2^\T = (b_{21},\cdots, b_{2w_2}), \\
  a_2(Z,X,M)^\T = \{ M^\T a_{2h}(Z,X)^\T, \cdots, M^\T a_{2w_2}(Z,X)^\T \}.
 \end{gather*}
\end{theorem}   
The $\mathrm{NIE}$ and $\mathrm{NDE}$ can be immediately identified when $ b_1$ and $\{ b_{2h} \}_{h=1}^{ w_2}$ are derived. We present a brief proof of the above theorem below. We rewrite the outcome as similarly as the main paper:
\begin{align*}
    Y &= a_1(Z,X;b_1) + \sum_{h=1}^{w_2}  b_{2h} a_{2h}(Z,X) M + \varphi^{\T} U + \eta \\
      &= a_1(Z,X;b_1) + \sum_{h=1}^{w_2}  b_{2h} a_{2h}(Z,X) M + \varphi^{\T} L + \psi,
\end{align*}
where $\psi = \eta +\varphi^\T U -b^\T(\Gamma U +\varepsilon)$. Then we show $\psi$ is indeed uncorrelated with $a_1(\cdot,\cdot,;b_1)$ and $\{ a_{2h}(Z,X) M \}_{h=1}^{w_2}$. Completely same as the proof of Theorem 1, $L$ must be uncorrelated with $M$ and $\check{M}$. Also, it is obvious that
$$
\mathrm{cov}\{ \psi, a_1(Z,X) \} = \mathrm{cov}\{ \eta + \varphi^\T U - b^\T (\Gamma U + \varepsilon), a_1(Z,X;b_1) \} = 0,
$$
which is due to the independence between $(Z,X)$ and $(U,\varepsilon,\eta)$. Also, we have
$$
\mathrm{cov}\{ \psi, a_{2h}(Z,X) \check{M} \} = E \{ \psi  \check{M}^\T a_{2h}(Z,X)^\T  \} = E \big( \psi \check{M}^T \big) E \{ a_{2h}(Z,X)^\T  \} = 0 \text{ for all }h=1,\cdots,w_2,
$$
where the last equality is due to the uncorrelated relation between $\psi$ and $\check{M}$. The rotation of $L$ also does not alter the conclusion, as illustrated similarly as the proof of Theorem 1. Thus, all above implies that $ b_1$ and $\{ b_{2h} \}_{h=1}^{ w_2}$ are both identifiable. Also, when the outcome model contains some unknown components, identification can still be achieved. Consider the following model:
\begin{align}
    Y &= d_1(Z) + d_2(X) + \sum_{i=1}^{w_2} b_{2h} a_{2h}(Z,X) M + \varphi^\T U + \eta, \label{modelforY_unknown} \\
    \Check{M} &= \Gamma U + \varepsilon, \label{modelforM_unknown}
\end{align}
where $d_1(\cdot)$ and $d_2(\cdot)$ are two unknown smooth functions, which can be expressed as the linear expansion of power series; that is, $d_1(Z) = \sum_{i=1}^{\infty} s_{1i} Z^i, d_2(X) = \sum_{i=1}^{\infty} s_{2i} X^i$ with two real sequences $\{ b_{1i} \}_{i=1}^{\infty}$ and $\{ b_{2i} \}_{i=1}^{\infty}$. Illustrations for other parameters and functions remain the same as all above. In this case, we summarize our identifiability results in the following theorem.

\begin{theorem} \label{unknow_case} 
     Under model~\eqref{modelforY_unknown}-\eqref{modelforM_unknown}, $ \{ d_1(\cdot), d_2(\cdot)  \}$ and $\{ b_{2h} \}_{h=1}^{ w_2}$ are identifiable if the following conditions hold:
	\begin{itemize}
		\item[(i)] after deleting any row of $\Gamma$, there remain two disjoint sub-matrices of full column rank;
		
		\item[(ii)] $L,a_2(Z,X,M),\{ Z^i \}_{i=1}^{\infty}$ and $ \{ X^i \}_{i=1}^{\infty}$ are linearly independent.
	\end{itemize}
\end{theorem}
The proof is straightforward and we omit it for simplicity. With the above identifiability, $\mathrm{NIE}$ and $\mathrm{NDE}$ can also be directly calculated.
\end{appendix}
\end{document}